\documentclass[final,5p,times]{elsarticle}

\usepackage{graphicx}
\usepackage{amssymb}

\usepackage{amsmath}

\usepackage{verbatim} 
\usepackage{float}

\usepackage{csquotes}
\usepackage{tabularx}
\usepackage{multirow}
\usepackage{color}
\usepackage[export]{adjustbox}

\usepackage[hyphens]{url}

\biboptions{authoryear}

\journal{Astronomy and Computing}

\begin{document}

\begin{frontmatter}


\title{CosmoHub: Interactive exploration and distribution of astronomical data on Hadoop}



\author[ciemat,pic]{P. Tallada\corref{tech_author}}
\ead{tallada@pic.es}
\author[ifae,pic]{J. Carretero\corref{sci_author}}
\author[ciemat,pic]{J. Casals}
\author[ifae,pic]{C. Acosta-Silva}
\author[ice,ieec]{S. Serrano}
\author[ciemat,pic]{M. Caubet}

\author[ice,ieec]{F.~J. Castander}
\author[uab]{E. C\'{e}sar}
\author[ice,ieec]{M. Crocce}
\author[ifae,pic]{M. Delfino}
\author[ifae,pic]{M. Eriksen}
\author[ice,ieec]{P. Fosalba}
\author[ice,ieec]{E. Gazta\~{n}aga}
\author[ifae,pic]{G. Merino}
\author[ifae,pic]{C. Neissner}
\author[bsc]{N. Tonello}

\cortext[tech_author]{Main technical author}
\cortext[sci_author]{Main scientific author}

\address[ciemat]{Centro de Investigaciones Energ\'eticas, Medioambientales y Tecnol\'ogicas (CIEMAT), Avenida Complutense 40, 28040 Madrid, Spain}
\address[ifae]{Institut de F\'isica d'Altes Energies (IFAE), The Barcelona Institute of Science and Technology, Campus UAB, 08193 Bellaterra (Barcelona), Spain}
\address[ice]{Institute of Space Sciences (ICE, CSIC), Campus UAB, Carrer de Can Magrans, s/n, 08193 Bellaterra (Barcelona), Spain}
\address[ieec]{Institut d'Estudis Espacials de Catalunya (IEEC), 08034 Barcelona, Spain}
\address[bsc]{Barcelona Supercomputing Center (BSC), C/ Jordi Girona 29, 08034 Barcelona, Spain}
\address[uab]{Universitat Aut\`{o}noma de Barcelona (UAB), 08193 Bellaterra (Barcelona), Spain}

\fntext[pic]{also at Port d'Informaci\'{o} Cient\'{i}fica (PIC), Campus UAB, C. Albareda s/n, 08193 Bellaterra (Barcelona), Spain }

\begin{abstract}

We present CosmoHub (\url{https://cosmohub.pic.es}), a web application based on Hadoop to perform interactive exploration and distribution of massive cosmological datasets.
Recent Cosmology seeks to unveil the nature of both dark matter and dark energy mapping the large-scale structure of the Universe, through the analysis of massive amounts of astronomical data, progressively increasing during the last (and future) decades with the digitization and automation of the experimental techniques.

CosmoHub, hosted and developed at the Port d'Informaci\'o  Cient\'ifica (PIC), provides support to a worldwide community of scientists, without requiring the end user to know any Structured Query Language (SQL). It is serving data of several large international collaborations such as the Euclid space mission, the Dark Energy Survey (DES), the Physics of the Accelerating Universe Survey (PAUS) and the Marenostrum Institut de Ci\`encies de l'Espai (MICE) numerical simulations.
While originally developed as a PostgreSQL relational database web frontend, this work describes the current version of CosmoHub, built on top of Apache Hive, which facilitates scalable reading, writing and managing huge datasets.
As CosmoHub's datasets are seldomly modified, Hive it is a better fit.

Over 60 TiB of catalogued information and $50 \times 10^9$ astronomical objects can be interactively explored using an integrated visualization tool which includes 1D histogram and 2D heatmap plots.
In our current implementation, online exploration of datasets of $10^9$ objects can be done in a timescale of tens of seconds. Users can also download customized subsets of data in standard formats generated in few minutes.

\end{abstract}


\end{frontmatter}


\section{Introduction}
\label{S:Introduction}

Experimental astronomy has entered in recent years into a new data regime, mainly due to the construction and development of ground \textemdash{}and space\textemdash{} based sky surveys\footnote{See \url{http://www.astro.ljmu.ac.uk/~ikb/research/galaxy-redshift-surveys.html} for a non-complete list of galaxy surveys} in the whole electromagnetic spectrum, from gamma rays and X-rays, ultraviolet, optical, and infrared to radio bands. This trend will increase with the next generation of projects, for example:
(i) the future 3.2 GigaPixel LSST camera \citep{LSST:2012} will take images every 30 seconds and the data rate will be about 15 terabytes per night\footnote{\url{https://www.lsst.org/about/dm}},
(ii) the complete Euclid survey \citep{Euclid:2012} represents hundreds of thousands images and several tens of petabytes of data; the final Euclid source catalog will contain about $10^{10}$ entries\footnote{\url{https://www.euclid-ec.org/}}.

A substantial part of the success of a scientific project can be measured by the impact its results have on the scientific community. Also, having powerful tools to facilitate exploration and distribution of data is key to boost their usage. 
With open science principles in mind, cosmology surveys are developing different solutions to share and distribute their data, including analysis tools.

One of the most successful and innovative galaxy surveys is the Sloan Digital Sky Survey (SDSS) \citep{York:2000}. The enormous success of this project is due to \textemdash{}besides the quality of the data\textemdash{} the fact that its results are fully public and easily accessible\footnote{\url{https://www.sdss.org/dr15/}}. They have put great effort into facilitating scientific exploitation by any user, regardless of their technical expertise (see \citet{2002cs........2013S} and \citet{2017AAS...22923615R}).

Most recent surveys have also created dedicated portals to manage access to their data releases. For example, the Dark Energy Survey \citep{DES:2005} has produced the DES Science Portal \citep{DESportal:2018}.
Future surveys like LSST are putting a tremendous effort into designing adequate tools to access and analyze the massive amounts of data they will generate \citep{juric2017lsst}.

We started developing CosmoHub in 2013, a web application for the interactive exploration and distribution of massive cosmological datasets. With its intuitive user interface, users with no Structured Query Language (SQL, \citet{Chamberlin1974}) knowledge could visualize and download customized subsets of the data. This first version used PostgreSQL to handle the data, following a similar approach as the one adopted by other projects (SDSS uses Microsoft SQL Server, while DES uses an Oracle Database). A few years later, we started struggling with performance issues due to the increasing amounts of data we were managing, and we decided to revisit our design choices.

This article describes CosmoHub as released in late 2016, powered by Apache Hive, a data warehouse solution based on Apache Hadoop which facilitates reading, writing and managing large datasets. First described in \cite{inproceedings}, CosmoHub is one of the earliest implementations of a storage and computing platform for cosmological datasets based on Apache Hive.

In the version described in this paper, over 30 TiB of catalogued information and $50 \times 10^9$ astronomical objects from a dozen different projects can be interactively explored using an integrated visualization tool which includes 1D histogram and 2D heatmap plots.
Interactive visualization of datasets of thousands of millions of objects can be done in less than a minute, and customized subsets can be generated in a timescale of minutes.

This paper is structured as follows: section \ref{S:CosmoHub} describes the main objectives of CosmoHub, our solution design and the evolution from the early prototypes, section \ref{S:Implementation} shows in detail the current implementation: the Hadoop platform, the backend and the frontend. Section \ref{S:Results} presents some results and use cases and finally section \ref{S:Conclusions} summarizes and concludes.

\section{CosmoHub}
\label{S:CosmoHub}

This section presents an overview of CosmoHub objectives, the technical implementation, and the historical development which led to the current form.

\subsection{Objectives}
\label{S:Objectives}

Most of the CosmoHub's objectives originated from our experience in designing and developing the PAU Survey data management (PAUdm) \citep{TONELLO2019171}, and from the interactions of the PAU Survey project with other peer projects such as MICE (\cite{10.1093/mnras/stv138mice}, \cite{10.1093/mnras/stv1708mice}, \cite{10.1093/mnras/stu2464mice}, \cite{10.1093/mnras/stu2402mice} and \cite{mice_growth}) and DES.
The following list defines the set of key requirements for CosmoHub.

\begin{itemize}

    \item \textit{Centralized data distribution}
    
    Having a unique point of data distribution enables having a single, authoritative version of the data, reducing the risk of duplicated, corrupted or deprecated replicas. A unique entry point also facilitates the enforcement of access controls and policies.
    
    Note that, relying on a common platform does not imply having a single point of failure. This platform can be configured in a high-availability setup where service is not disrupted by eventual failures of its individual components (see Section \ref{fig:hardware}).

    \item \textit{Easy to use}
    
    Usability is also a key for the success of this kind of platform. The easier it is to use, the more users it will engage and, therefore, the data published on it will reach a wider audience, increasing their impact on the scientific community.
  
    Interfaces should be clean and simple enough such that any user may use the service without prior training. In detail, the following two issues should be addressed:
    
    \begin{itemize}
    
    \item \textit{No Structured Query Language (SQL) knowledge must be required}
    
    In a data distribution service, SQL is the most common interface for interacting with the data. SQL is a declarative language that provides a set of constructs to select, project, filter and retrieve subsets of information from a database.
    As an industry standard, most (if not all) vendors of data warehouses offer a SQL interface to interact with their own services.
    
    Exposing an SQL interface is problematic for at least two reasons: First, while SQL knowledge is common in technical circles, many scientific users are unfamiliar with SQL. And second, SQL is an industry standard, but has different vendor implementations that deviate from official specifications. These differences originate from adding complementary features or because their implementation predates the official specification.
    This means that even users with training on data warehouses might encounter problems because the SQL they know does not match the exact flavour used in a given solution.
    
    \item \textit{"Common" file formats}
    
    The astronomical community has grown used to a standard set of formats for data interchange. 
    In particular, the approach of the Virtual Observatory (VO) led by the International Virtual Observatory Alliance (IVOA\footnote{\url{http://ivoa.net}}) is of most importance. Consequentially, specific tools for managing, processing and visualizing data stored in these formats have been developed and are widely used. Therefore, we must support those formats to enable our users to keep using the tools they already master and make the interoperability with them as straightforward as possible.
    
    \end{itemize}
  
    This usability objective should not only be considered for end users, but also for system administrators. They should be able to deploy and keep the service operational, avoiding that eventual hiccups pose a threat to service availability.

    \item \textit{Custom datasets}
    
    One of the main challenges of managing large datasets is to be able to efficiently generate small customized subsets fitting the scientists' needs.
    
    Allowing the generation and download of custom subsets enables users to minimize data storage and transfer costs. At the same time, processing costs are also reduced as the selection and filtering part is offloaded onto the service, which has plenty of resources optimized for this task.
    
    As a downside, the service must cope with the additional storage required for all (potentially overlapping) subsets that the users have requested, including some means to eventually expire or purge them.
    
    \item \textit{Quick exploration}
    
    The ability to create custom datasets is not useful if the user does not know the exact criteria to specify them or is not confident enough on the properties of the resulting subset. Having some functionality for the user to interactively explore and preview the results of subset generation is very helpful.
    
    This quick exploration tool can be offered in different ways, for example row sampling or simple visualizations such as scatter plots and histograms. Once the users are certain that the subset matches their expectations, they can proceed with the download, as needed.
    
\end{itemize}

\subsection{Solution design}
\label{S:Solution design}

In this section we describe the main aspects of the design of CosmoHub.

\begin{itemize}

    \item \textit{Target audience}
    
    Prospective users of CosmoHub are the thousands of scientists around the globe collaborating in astronomical projects that manage and/or produce large amounts of catalogued data. Most of these data end up released to the public and, if a replica is available in CosmoHub, any registered user may access, explore and download it.
    
    Furthermore, these projects usually have private datasets for internal processes, such as release validation or calibration, which are only available to project members.
    
    \begin{figure}
    \centering
    \includegraphics[width=1.0\columnwidth]{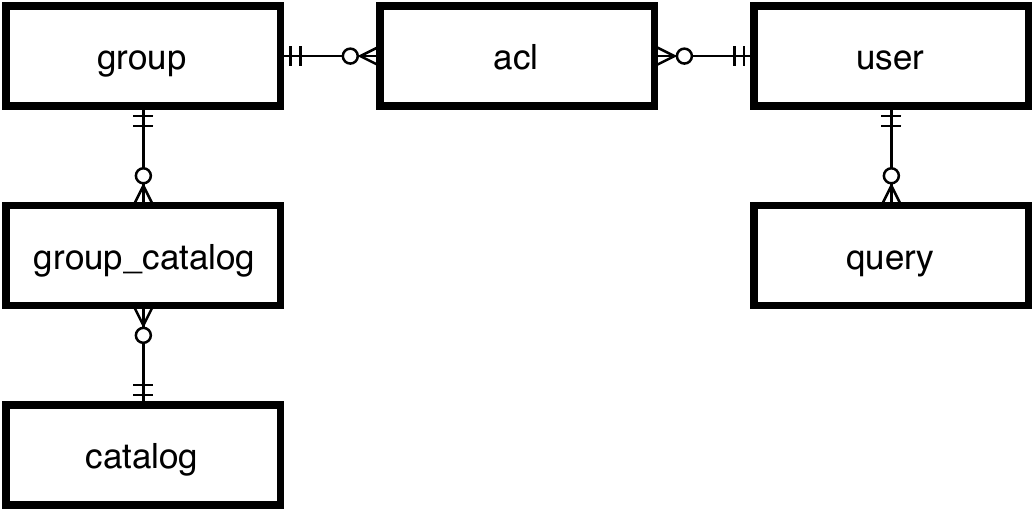}
    \caption{
    CosmoHub's data model, which stores information about its catalogs, groups, users and access control lists (acl), among other metadata.
    }
    \label{fig:data model}
    \end{figure}
    
    \item \textit{Metadata database}
    
    CosmoHub provides access to a collection of catalogs or astronomical datasets hosted in our data warehouse. Each catalog is defined by a name, a short description and a summary of its characteristics. Each catalog is mapped to a single table in our data warehouse containing a set of columns, which are described by its data type and a short description.
    
    Each catalog belongs to one or several projects or research groups. Users can access all data in CosmoHub associated to the groups they are member of. Users request access to the groups when registering in the service. Each group has a set of users with special privileges who are in charge of validating and granting the corresponding membership requests. Access to specific groups or projects can be updated at any time through a web interface. Users receive an email notification whenever their privileges change. A catalog may also be public and, as such, not require any specific membership to access it.
    
    Figure \ref{fig:data model} shows the relational data model of all this metadata. In particular, in addition to storing the information about catalogs, groups and users, it also includes the relationships between them such as project membership or group administration privileges.
    
    \item \textit{Web interface}
    
    CosmoHub is designed as a web application.
    This solution only needs a web browser, a requirement that usually comes preinstalled on any computer. 
    Moreover, this also allows to reuse all the user experience of the web semantics and graphical metaphors that most users are already used to.
    
    \item \textit{Guided process}
    
    In order to facilitate the usage of the platform to those users without SQL knowledge, the custom subset building interface has been designed as a guided series of steps that can be followed very easily even by the most inexperienced user.
    
    The web interface guides users through a sequence of steps, allowing them to select catalogs they are interested in, then the columns they need, adding filtering criteria (if needed) and choosing the download format.
    
    They can also plot and preview the dataset they are building with the integrated plotting tool, also implemented with intuitive and easy to use web forms to configure each type of plot (see section \ref{S:Frontend} for details).
    
    \item \textit{SQL expert mode}
    
    CosmoHub also offers the possibility to unleash the full power and capabilities of SQL. The "expert" mode allows to write an SQL query directly and passes it to the underlying database for its execution. This feature allows more experienced users to define additional computed columns using standard functions and operators, specify arbitrary groupings or even perform joins. The latter is very interesting because it allows matching and combining data from several different catalogs. 
    
    Furthermore, these capabilities can be extended by implementing additional user defined functions. For instance, CosmoHub includes functions for dealing with the HEALPix\footnote{\url{http://healpix.sourceforge.net/}} \citep{2005ApJ...622..759G} pixelization scheme.

    \item \textit{(Simple) visualization}
    
    Users may get a quick insight on the data they are selecting by using the four integrated visualizations that CosmoHub offers:
    
    \begin{itemize}
        \item Table overview shows 20 rows of the subset.
        \item Scatter plot, to visualize trends and relations between different columns. It also supports plotting different series of data, but is limited to plot only 10k points, so users only see partial results/behaviour of the full subset. Therefore, the plot may not be representative of the subset as a whole.
        \item 1D histograms and
        \item 2D heatmaps, both with automatic hints on column names, and bin ranges.
    \end{itemize}
    
    1D histograms and 2D heatmaps are aggregated plots implemented on the backend. They get the full picture of the custom subset. For performance reasons they are limited to less than 10k uniform bins, which is by far enough for most applications.
    
    \item \textit{Batch custom subsets}
    
    When users finish exploring a subset, they can select a download format and request its generation and delivery. Among the formats, special attention must be put in supporting Flexible Image Transport System (FITS) \citep{wells1979fits}, which is one of the most popular data formats in astronomy.
    
    The custom subset is built in the background by the underlying database engine. When the custom subset is ready and stored, an email will be sent to the user with a link that they must follow in order to start the download.

\end{itemize}

\subsection{Early prototypes}
\label{S:Early prototypes}

CosmoHub history can be traced back to the beginnings of the Physics of the Accelerating Universe Survey (PAUS) project in 2013. At that time, PAUS had started to produce its first simulated data that needed to be distributed to its collaborators. In order to facilitate the distribution, a pilot web interface called "PAUdm Simulations Portal" was commissioned and integrated into the official PAUdm operations web. This prototype offered access to the PAUdm database hosted in a PostgreSQL server.

The amount of data stored in the prototype grew substantially. Most of it came from external data used in PAUS pipelines, such as SDSS star catalogs and MICE mock galaxy catalogs. In time, MICE started ingesting more of its own data in the same database, in order to be able to use the web interface. Through several iterations we ended up implementing some dataset exploration features.

From the beginning, CosmoHub was designed in a way that no specific technical knowledge was required in order to exploit its functionalities. In particular, users were already able to formulate queries without any SQL knowledge through a guided process. Also, users were able to directly download Value-Added-Data (prebuilt catalogs or other information needed to analyze the data, such as filter curves or dust maps), they could visualize general data trends using simple plots and, of course, download custom subsets which were created asynchronously in the computing facilities at PIC.

After a year or so, the performance of the database server \textemdash{}designed specifically to host only PAUS data\textemdash{} started to suffer. The amount of data hosted kept growing, mainly due to catalogs ingested from external projects, the storage space became tight and response times degraded. 

In this situation, the first affected feature were the interactive plots, which at that time were limited to 10k rows and queries taking less than 2 minutes to complete. With the increasing size of the catalogs, most queries did not fulfill the response time requirements even using custom indexes.

We ended up migrating to another instance of PostgreSQL database in a separate server, with much more storage space and similar processing power. In that way, we mitigated two problems: the limited storage available and the competition of computing resources with the main PAUdm processing pipelines.

Nevertheless, the problem of solving the long response times was hard to tackle. In our experience, traditional relational databases such as PostgreSQL can deal with huge datasets, as long as you deal with them in small chunks. 
But when the requested data is above a certain threshold, the PostgreSQL query optimizer does not use indexes as it is not efficient anymore. That is the main reason why most of the queries ended up performing a sequential scan of the entire table, resulting in response times that can go from hours to even days. Last but not least, modifying the schema and removing large amounts of rows were extremely inefficient operations.

For these reasons we realized that PostgreSQL was not the right tool for CosmoHub's data workflow and we explored different possibilities (see section \ref{S:Hadoop}) to solve the problem. In the end we decided to use Apache Hive, a data warehouse based on Hadoop.

\section{CosmoHub implementation}
\label{S:Implementation}

This section introduces the implementation details of CosmoHub's main components.
First, we discuss the configuration of the Hadoop platform that hosts the data and supports the processes that implement the available services, e.g. interactive plots and custom catalogs. Then, we describe how input catalogs are ingested and stored, how SQL queries are treated, which are the output formats available to users, and how contents are generated. Next, we introduce the design and implementation using Python of the Representational State Transfer (REST) API, which implements the available operations to the users, hiding the particular details of the underlying Hadoop platform. Finally, we describe the implementation of the web interface and workflows that allow users to interact with CosmoHub.   

\subsection{Hadoop Platform}
\label{S:Hadoop}

\begin{figure*}
\centering
\includegraphics[width=1\linewidth]{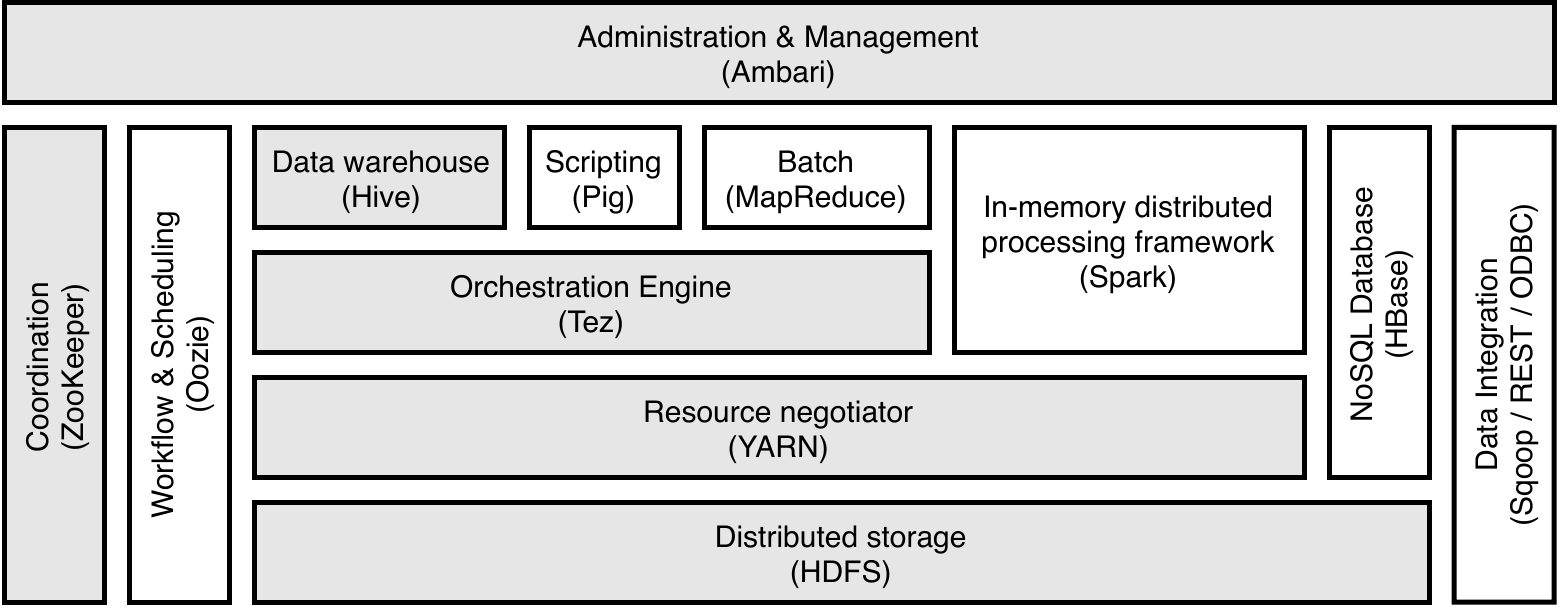}
\caption{
Hadoop layered ecosystem, showing how multiple components stack on top of others to provide a broad set of features and services. In grey, the components used for CosmoHub,
}
\label{fig:hadoop stack}
\end{figure*}

\subsubsection*{Assessing alternatives}
From the experience gained from prototypes, we knew that choosing the right data storage and processing platform was fundamental to achieve our objectives. Therefore, we researched and tested several alternatives that we thought could be useful. There are multiple solutions in the market to handle large structured datasets in a manner that is scalable and has good performance, such as NoSQL (or non-relational) databases and distributed relational databases. We only took into consideration the open source alternatives due to technical and economic reasons.

We knew from the beginning that one of the key requirements of CosmoHub would be the ability to hold multiple large datasets, and to be able to analyze, compare and crossmatch them. 
The particular architecture of NoSQL solutions such as HBase\footnote{\url{https://hbase.apache.org}}, Cassandra\footnote{\url{https://cassandra.apache.org/}}, MongoDB\footnote{\url{https://www.mongodb.com/}} or Redis\footnote{\url{https://redis.io}}) normally does not allow for efficient joins between datasets. Furthermore, each of them implement a different language for interacting with the data, with partial SQL support, at best.
For these reasons, we decided to discard the NoSQL solutions.

For the distributed relational databases, we studied two approaches. First, we tested clustered implementations of traditional databases, such as Postgres-XL and Greenplum. These solutions rely on sharding and replicating the datasets onto multiple nodes in a computer cluster. Queries are then split and routed to the proper nodes, which execute them assisted by a central coordinator. Due to the reliance on partitions and indexes, this kind of solutions are optimal for scaling out large datasets.

These kind of solutions are mostly engineered to the CRUD (Create, Retrieve, Update and Delete) paradigm they have inherited, where each operation usually involves a small subset of the total number of rows. In contrast, in the typical CosmoHub workflow, datasets are ingested and deleted always as a whole, never updated, and usually retrieved in large subsets, or as aggregations of large subsets. In addition to the critical differences in data workflow design, we found that it was not straightforward to implement or integrate new data formats on these solutions.

Next, we tested solutions based on the Hadoop platform, 
such as Apache Hive \citep{Thusoo:2009:HWS:1687553.1687609} and Apache Impala \citep{bittorf2015impala}. Hive is an open-source data warehouse which has gained a lot of momentum since 2013, mostly thanks to the Hortonworks\footnote{\url{https://hortonworks.com/}} Stinger\footnote{\url{https://es.hortonworks.com/blog/100x-faster-hive/}} and Stinger.next\footnote{\url{https://es.hortonworks.com/blog/stinger-next-enterprise-sql-hadoop-scale-apache-hive/}} initiatives. Impala is a massively parallel processing (MPP) SQL query engine for data stored in Hadoop, and its most important contributor is Cloudera\footnote{\url{https://www.cloudera.com/}}.

When evaluating these alternatives, we found that Impala timings were inconsistent, and in some cases, the results were incorrect.
More up to date correctness studies have replicated the same findings\footnote{\url{https://mr3.postech.ac.kr/blog/2018/10/30/performance-evaluation-0.4/}} \footnote{\url{https://mr3.postech.ac.kr/blog/2019/03/22/performance-evaluation-0.6/}} \footnote{\url{https://mr3.postech.ac.kr/blog/2019/06/26/correctness-hivemr3-presto-impala/}}. Furthermore, the administration tools from the Cloudera Hadoop distribution were not free, compared to the Hortonworks open source ones. Consequently, we decided on a solution based on Apache Hive on top of Hortonworks due to its stability, extensibility, comprehensive documentation and availability of free administration tools.

\subsubsection*{Software stack}
Apache Hive is one of the multiple components in the Hadoop ecosystem. Figure \ref{fig:hadoop stack} displays a typical Hadoop architecture showing several of those components layered in a stack. CosmoHub heavily relies on several of these components, specially on Apache Hive, which is a data warehouse software that facilitates reading, writing, and managing large structured datasets located in distributed storage. From the administrator's point of view, migrating from the previous setup based on PostgreSQL to Apache Hive is easy as both have the same interface (SQL) with the data.

Apache Hive sits on top of Apache Tez \citep{Saha:2015:ATU:2723372.2742790}, an application framework which allows the execution of complex directed-acyclic-graphs (DAG) of tasks for processing data. The scaffolding provided by Tez allows the orchestration and optimization of Hive tasks, even at runtime, boosting its performance.

The computing needs of every Hadoop platform are delivered by Yet Another Resource Negotiator (YARN) \citep{yarn}. YARN enables the execution of arbitrary tasks on top of containers executed in cluster nodes. Their resources are delimited and isolated, as to not interfere or starve each other. Resources are managed using scheduling queues, where each user and group can get a fair share of resources as per configuration.

The keystone of the Hadoop platform, the Hadoop Distributed File System (HDFS) \citep{hdfs5496972} lays at the base of the stack. HDFS is a high performance distributed filesystem that makes use of the storage of the nodes in a Hadoop cluster, merging it into a single name-space for increased capacity and performance. Concisely, it works by splitting the files in fixed-size blocks and then replicating those blocks between the available nodes in the cluster. This architecture allows very good resilience and scalability.

Additional components take care of security, user authentication and authorization, as well as the administration and configuration of the different components. The market offers several distributions that include most of those components in the form of self-contained packages that facilitate an easy installation and configuration and, in addition, commercial support. In our case, all these layers are provided by the Hortonworks Data Platform (HDP) software solution. The currently installed version is HDP 2.6.5.

\begin{figure*}
\centering
\includegraphics[width=1\linewidth]{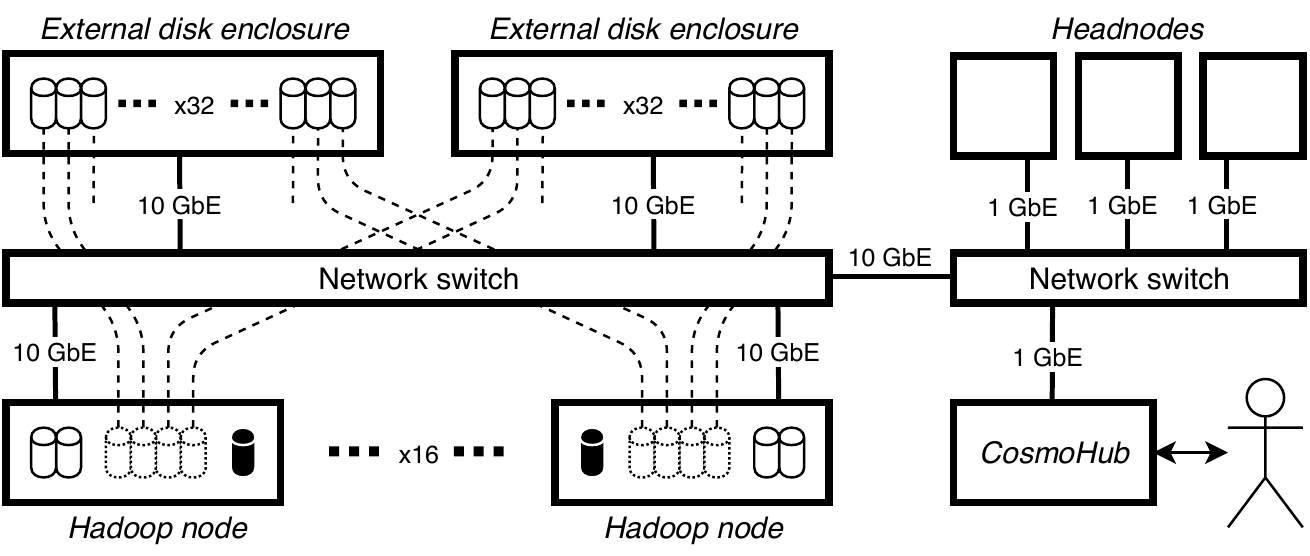}
\caption{
Hadoop cluster architecture, showing the configuration of the external disks and their logical links, the physically separated network for the nodes and external disk enclosures and the connection with the core network along with the headnodes and CosmoHub server.
}
\label{fig:hardware}
\end{figure*}

\subsubsection*{Hardware structure}
Figure \ref{fig:hardware} shows the hardware architecture of the Hadoop platform used for CosmoHub. The current cluster is composed of 16 compute nodes and 3 head nodes. The 16 compute nodes are grouped in 4 dual-twin servers. Each node is equipped with two 14-core CPUs, 128~GiB of RAM, two 6~TB SATA disks and a single 960~GB SSD disk. The SATA drives are configured with a small RAID-1 partition to host the OS, and the rest of space is used for HDFS storage (without RAID). The SSD drive is devoted to cache intermediate files in order to boost the efficiency of shuffle operations \citep{Kambatla:2014:TMP:2717491.2717499}, such as CosmoHub joins. The compute nodes are located at our innovative oil-submersion cooling facility, which is very power efficient \citep{acin}.

Because of the dense form factor used for the compute nodes, there is not much space for expansion. Thus, in order to expand the storage capacity, we provisioned two external disk enclosures, with 36 SATA disks of 2~TB. Each compute node mounts 4 of those disks using iSCSI, two from each enclosure, accounting for 32 of the 36 drives in each enclosure. The remaining 4 drives are left as spare. The iSCSI disks appear as if they were local disks and are used for HDFS storage. 

All in all, the compute nodes sum 448 cores, 2~TiB of RAM, 192~TiB of local SATA storage space and 128~TiB of external iSCSI storage space. After taking into account the CPU and RAM reserved for Hadoop services in the compute nodes, and the overhead that replicas introduce in HDFS storage, the actual resources available for processing are approximately $\sim$400 cores, 1.8~TiB RAM, $\sim$60~TiB of local SATA storage and $\sim$40 TiB of external iSCSI storage space.

Each compute node and disk enclosure has a 10~GbE link to the same network switch. This switch is kept separated from the rest of the network infrastructure so that traffic peaks or saturation do not affect other critical services at PIC. An additional 10~GbE link connects this switch to the rest of the PIC network, where the head nodes and the CosmoHub application server are located. 

The head nodes are equipped with two 8-core CPUs, 32~GiB of RAM and 2~TB of local storage. They are configured in a high-availability setup. In case of a failure, the service running in another head node is able to promote itself to master and keep the service functioning.

\subsection{Data workflow}
\label{S:Workflow}

CosmoHub hosts multiple datasets of different projects, origins, cardinalities and complexities, and all of them are stored on HDFS.
Most of the time, incoming raw data is published in some kind of ASCII\footnote{\url{https://tools.ietf.org/html/rfc20}} format, like CSV. The main problem with this kind of formats is that they are not able to describe the corresponding data types in a native way. In some cases, the data files are augmented with additional headers or they come with attached documentation of the types and meaning of the different columns in the dataset. However, alternative binary formats \textemdash{}such as FITS\textemdash{} are also common, and lately others such as Hierarchical Data Format 5 (HDF5) \citep{hdf5} have been gaining traction. A decisive advantage of these binary formats is that they can store the machine representation of the values and they carry along detailed metadata, including the description of data types and columns. Therefore, they are preferred in order to preserve as much information as possible when ingesting a dataset into CosmoHub. 

\subsubsection*{Raw data}

The original upstream data, in whatever raw format is provided, is copied into HDFS and then converted into a native Hive format that is suitable for efficient query processing.
Optimized Row Columnar (ORC\footnote{\url{https://orc.apache.org/specification/ORCv1/}}) and Parquet\footnote{\url{https://parquet.apache.org/}} are two contending formats in this area.

Features like columnar-based structure, push-down predicate (PPD) capabilities, column statistics or bloom filters are very useful for query efficiency, as they enable to skip entire sections of rows that do not contain data of interest.
While both are very similar and support nearly the same set of features, we decided to use ORC because it is the one that is recommended by Hortonworks, the chosen Hadoop platform distribution for CosmoHub.

\subsubsection*{Interactive exploration}
\label{S:Exploration}

Queries intended for visualization purposes should finish in a few seconds to satisfy the interactivity requirement.
As transferring large amounts of data to the browser for plotting is unfeasible, we use histograms and heatmaps to visualize large datasets.
CosmoHub rewrites the queries that feed the histogram and heatmap plots in order to pre-aggregate the results and deliver to the browser only the data points to be plotted. This minimizes network traffic and lessens the load on the client.

Furthermore, in order to speed up even more the execution of interactive queries, two resources queues have been set up in YARN. One queue is for batch tasks such as generating custom catalogs or other non-CosmoHub related jobs, while the other is devoted solely to the execution of interactive queries. This last queue has absolute priority over the resources and can even preempt resources from the other queue if needed in order to ensure the fastest execution time possible.

We cannot make any guarantees regarding the response times, as the nature of the interaction between the users and CosmoHub is very diverse and our resources are limited. Nevertheless, the results (see section \ref{S:Quantitative analysis}) show that, in nearly all cases, the interactivity requirement is achieved.

\subsubsection*{Custom catalogs}

Custom catalogs are generated from an SQL statement where the output is written concurrently as a set of files, each one of them created by a different cluster node. WebHCat\footnote{\url{https://cwiki.apache.org/confluence/display/Hive/WebHCat}} is used to orchestrate the execution of this SQL statement. In the previous version of CosmoHub, the catalogs were offered as a single file download. In order to have a smooth transition, we wanted to keep providing the same download interface.

Therefore, the download formats that we could support would need to have a fast and simple way to merge the set of individual files into a single stream. Taking into account this restrictions, CosmoHub implements three different download formats: BZip2-compressed Comma-Separated Values (CSV\footnote{\url{https://tools.ietf.org/html/rfc4180}}), FITS and Advanced Scientific Data Format (ASDF) \citep{greenfield2015asdf}.

\subsubsection*{Data formats}

CSV is a very well known ASCII format, broadly supported among all kinds of programming languages, toolkits and specialized software. However, it is also very bulky. Compression can greatly reduce its footprint, but not all compression codecs allow merging different files or streams into a single one. Hadoop supports different compression codecs, including GZip\footnote{\url{https://tools.ietf.org/html/rfc1951}} \footnote{\url{https://tools.ietf.org/html/rfc1952}}, BZip2 and XZ\footnote{\url{https://tukaani.org/xz/}}. GZip is fast, provides a decent compression factor, but does not support the concatenation of gzip-compressed streams. BZip2 and XZ both provide great compression factors at an increased computational cost, and both support the concatenation of its compressed streams.

We decided to use BZip2 for its stability and broad support compared to XZ. It is worth noticing that during our initial tests, we also found some issues\footnote{\url{https://github.com/yongtang/hadoop-xz/issues/9}} in the Hadoop XZ library.

FITS is an stable and mature format that has become the de-facto standard for astronomy data interchange. First defined in 1981, it has undergone several revisions (latest is v4.0 released in August 2018). However, it still carries on several limitations (see \citet{THOMAS2015133}) originating from the ancient hardware it was designed for.

In order to implement FITS as a download format in CosmoHub, the main limitation for us is its lack of streaming support. Without entering into much detail, FITS files are divided in headers and data sections. In our particular case, the FITS files used for storing tabular data define the schema of each row in the header, along with the number of rows. The row length depends on the schema and is equal for all rows. Both the header and data sections are null-padded to a multiple of 2880 bytes and saved to disk. The problem is that the number of rows that must be stored in the header is not determined until the query has finished. By then, the output has already been generated and HDFS does not allow random-access writes to a file, only appends. 

To circumvent this problem, we developed a custom Hive output format\footnote{\url{https://github.com/ptallada/recarrayserde}} that stores just the data section (without padding). We can compute the number of generated rows as the result of dividing the total size of the output by the row length once all outputs have been written, given that the row length is fixed. Once we know the number of rows, streaming the results back to the user is easy. We generate the header on-the-fly from the query metadata stored in the database, we pad it as a multiple of 2880 bytes and we serve it to the browser, appending after it all the output files the query has produced. At the end, we pad also the last output. The resulting downloaded file is a perfectly compliant FITS file.

ASDF is also supported as download format in CosmoHub. It has been designed as the successor of FITS, claiming to remove most of its problems. ASDF files are divided in several sections but, for our purposes, we are only interested in the tree and the data sections. The schema is defined in the tree section and is serialized as a YAML Ain't Markup Language (YAML) \citep{ben2009yaml} document.
The rows are stored as binary blocks in the data section.

In order to facilitate interoperability between FITS and ASDF, the block binary format is compatible with FITS. This fact made it very easy to add ASDF support in CosmoHub. As the rows are already written in FITS format, we just need to generate the ASDF header and append all the outputs without any padding.

\subsection{Python backend}
\label{S:Backend}

CosmoHub allows users to process large datasets to obtain additional information. In order to decouple the results of those interactions from the way they are presented, all operations are carried out through a set of API calls. In particular, all actions available to the user are implemented in an API that follows the REST \citep{Fielding:2000:ASD:932295} paradigm. Operations are grouped in several endpoints depending on the data model entity they act on (see Figure \ref{fig:data model}). Database access is proxied through an Object Relational Mapper (ORM) layer in order not to tie the implementation of each action to the data model specifics, enabling data model evolution. 
All actions return JSON\footnote{\url{https://tools.ietf.org/html/rfc7159}} responses, which are consumed by the web frontend. A complete list of endpoints and their description is available in Table \ref{table:endpoints}.

\begin{table*}
\centering
\renewcommand{\arraystretch}{1.3}
\begin{tabularx}{\textwidth}{|l l X|} 
  \hline
  URL & Method & Description \\ 
  \hline \hline
  \multirow{4}{*}{/user} & GET & Retrieve current user profile \\ 
  & PATCH & Update profile data (i.e. email, password) \\ 
  & POST & Register a new user \\ 
  & DELETE & Remove a user account \\ 
  \hline
  /groups & GET & Retrieve the list of groups \\
  \hline
  \multirow{2}{*}{/acls} & GET & Retrieve the list of users and their memberships \\
  & PATCH & Modify a user's membership \\ 
  \hline
  /catalogs & GET & Retrieve the list of catalogs accessible to the current user \\
  /catalogs/\{id\} & GET & Retrieve detailed information of a catalog \\
  /catalogs/syntax & GET & Perform an SQL syntax check \\
  \hline
  /downloads/datasets/\{id\}/readme & GET & Download the \textit{README} file for a dataset  \\
  /downloads/files/\{id\}/readme & GET & Download the \textit{README} file for a value-added file \\
  /downloads/files/\{id\}/contents & GET & Download the contents of a value-added file \\
  /downloads/queries/\{id\}/results & GET & Download the output of a custom catalog \\
  \hline
  \multirow{2}{*}{/queries} & GET & Retrieve the list of custom catalogs for the current user \\
  & POST & Request the generation of a custom catalog \\
  /queries/\{id\}/cancel & POST & Abort the generation of a custom catalog \\
  /queries/\{id\}/done & GET & Callback to notify the completion of a custom catalog \\
  \hline
  /contact & POST & Send a message to the CosmoHub Team \\
  \hline
\end{tabularx}
\caption{
List of REST API endpoints, grouped by entity. For each one, its URL pattern, the HTTP method and a brief description is shown.
}
\label{table:endpoints}
\end{table*}

Most catalogs in CosmoHub belong to a single project although, in some special cases, they can be associated with several projects. Only users which are members of those projects are able to access their corresponding data. In order to prevent unauthorized uses of CosmoHub, all requests are authenticated and the user privileges are checked against the database. The API accepts two authentication methods, basic and token.

With HTTP Basic authentication\footnote{\url{https://tools.ietf.org/html/rfc7617}}, each request must include a username and password combination. This information is looked up in the user database and, if no match is found, the request is denied. The main inconvenience with this mechanism is that each request requires a round-trip to the database. In order to soften the load on the database, a JSON Web Token\footnote{\url{https://tools.ietf.org/html/rfc7519}} (JWT) is attached to every response. This token contains signed information about the authenticated user and, when supplied on future requests, it allows the backend to verify the identity of the user without any database involvement.

Regarding the interactive exploration feature of CosmoHub, as Hive queries are potentially executed on all nodes in the Hadoop platform, a full table scan usually takes about a minute. Combined with sampling, results can be obtained even faster. This performance allowed us to implement interactive exploration of large datasets using histograms and heatmaps.
As also mentioned in \ref{S:Exploration}, data for these plots is pre-aggregated on the Hive side using a specially constructed query and only the data points to be plotted are sent to the browser.

Additionally, in order to provide some feedback to the user during the execution of interactive queries, an extension\footnote{\url{https://github.com/dropbox/PyHive/pull/136}} was developed for the Python DB-API interface for Hive (PyHive) in order to extract the progress of an ongoing query. This information is relayed using a websocket\footnote{\url{https://tools.ietf.org/html/rfc6455}} connection, which enables bidirectional communication between the browser and the backend.
Through this channel, users receive periodic progress updates about a query and can also request its cancellation.

CosmoHub is deployed in three different instances corresponding to the production, pre-production and test environments. Each one of them runs on separate identical virtual machines, with 4 cores, 4 GiB of RAM and 10 GiB of storage each. Only the production instance is accessible to the outside through \url{https://cosmohub.pic.es}.

The main software components used for building the backend stack are Flask\footnote{\url{http://flask.pocoo.org/}} as the Python Web Server Gateway Interface (WSGI\footnote{\url{https://www.python.org/dev/peps/pep-3333/}}) framework, Flask-RESTful\footnote{\url{https://flask-restful.readthedocs.io}} as the REST framework and gevent\footnote{\url{http://www.gevent.org/}} as the coroutine networking library. This stack runs on top of uWSGI\footnote{\url{https://uwsgi-docs.readthedocs.io}} behind an NGINX\footnote{\url{https://www.nginx.com/}} proxy. 

For the data access layer, SQLAlchemy\footnote{\url{https://www.sqlalchemy.org/}} is used as the ORM component and the combination of psycopg\footnote{\url{http://initd.org/psycopg/}} and psycogreen\footnote{\url{https://bitbucket.org/dvarrazzo/psycogreen}} are used as the PostgreSQL driver and coroutine adapter library, respectively.
Finally, astropy\footnote{\url{https://www.astropy.org/}} is used to implement FITS as a download format and the ASDF\footnote{\url{https://asdf-standard.readthedocs.io}} python library to implement ASDF format.

\subsection{Web frontend}
\label{S:Frontend}

The web interface's main objective is to enable the user to access all of CosmoHub's features which, as described in the previous section, are available through a set of REST endpoints.

Usability is a strong requirement, as the interface should be intuitive enough so that any user can interact with it with no prior training. Special care should be taken to follow and exploit characteristic web semantics \textemdash{}such as forms, hyperlinks or scrolling, among others\textemdash{} to aid the user at every step. In the end, designing a clean and simple interface, preferably self-explanatory, is key to permitting open science.

CosmoHub's frontend has been developed using modern, widely-supported, community technologies, such as AngularJS\footnote{\url{https://angularjs.org/}}, Bootstrap\footnote{\url{https://getbootstrap.com/}}, WebSockets\footnote{\url{https://tools.ietf.org/html/rfc6455}}, Plot.ly\footnote{\url{https://plot.ly/}} and Wordpress\footnote{\url{https://wordpress.com/}}.

AngularJS is a Javascript framework, developed and maintained by Google\footnote{\url{https://google.com}}, that extends HTML to implement Model-View-Controller capabilities into web browsers, making user interactions dynamic, faster and more fluid. It has a large collection of official and third-party plugins and is specially designed to interact with API based applications, such as CosmoHub. Being open-source, well maintained and with a broad community of users and developers are key aspects for choosing it as the base of the frontend.

Bootstrap is an open-source HTML, JavaScript and Cascading Style Sheets (CSS) library created and maintained by Twitter\footnote{\url{https://twitter.com}} for responsive web design. It is one of the most used styling frameworks in the entire web development community, providing users with clear and coherent interfaces. It comes with a handy and easy to configure column-based layout, plus some predefined style elements. These features are extensively used in CosmoHub.

WebSockets is a technology for bi-directional communication between web browsers and web servers.
This technique involves \textit{upgrading} a stateless HTTP request into a persistent TCP connection that can be subsequently used to transfer information from both parties. Some features such as real-time progress monitoring require websockets to work properly.

Plot.ly is an open-source plotting library based on the widely used D3.js\footnote{\url{https://d3js.org/}} web visualization framework. It greatly simplifies the programming needed to implement all sorts of charts and dashboards, such as those used for interactive exploration.

Finally, WordPress is one of the most used content management systems (CMS). Although it is mostly associated with blogs, CosmoHub uses it as a backend for editing the content of dynamic sections, such as the news feed.

\begin{figure}
\centering
\includegraphics[width=1\columnwidth,frame]{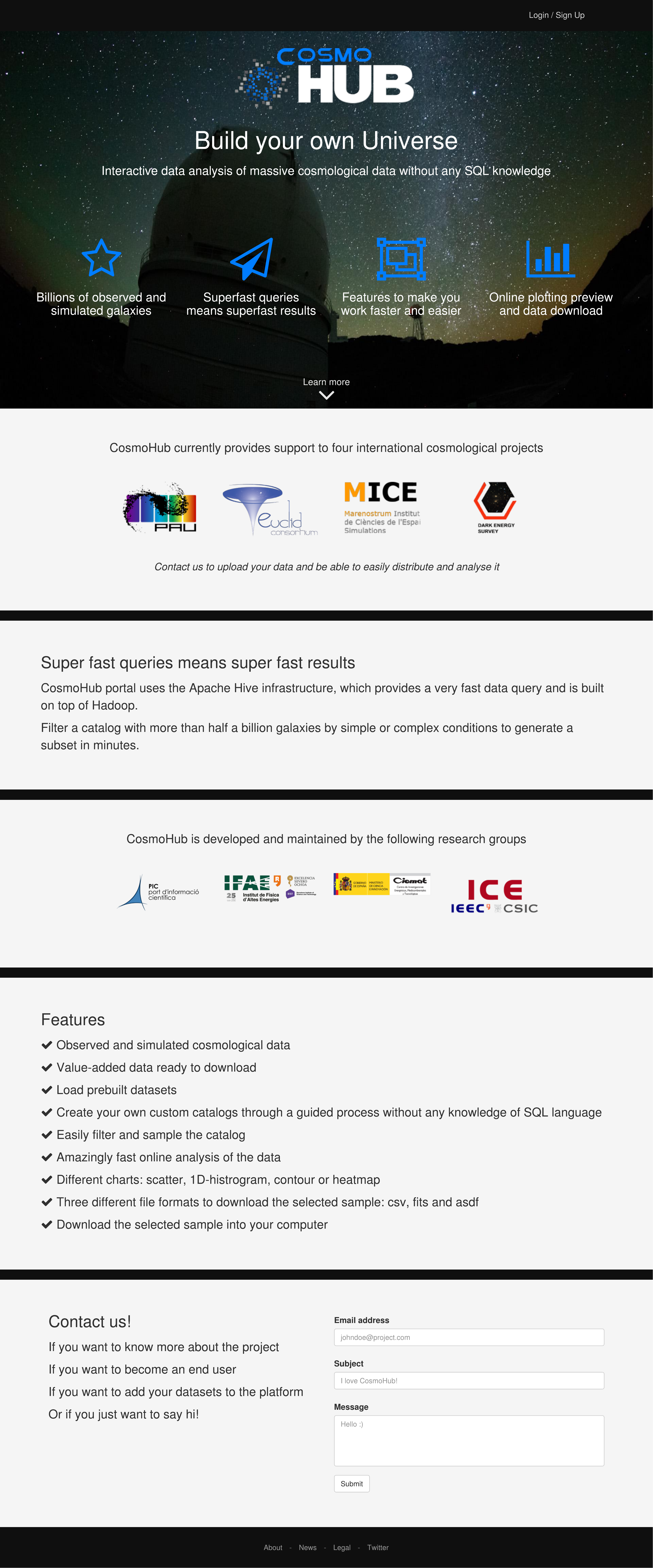}
\caption{
Initial page, showcasing the main projects and features. Note that some content has been edited for presentation purposes.
}
\label{fig:front page}
\end{figure}

When a user visits CosmoHub (\url{https://cosmohub.pic.es}), it is presented with the initial page shown in Figure \ref{fig:front page}. This front page describes its goals, showcases its main features and holds references to the rest of public contents: about page, news feed, terms of use and a link to the Twitter profile. All these sections can be accessed without authentication.

\subsubsection*{User management}

\begin{figure*}
\centering
\includegraphics[width=1\linewidth,frame]{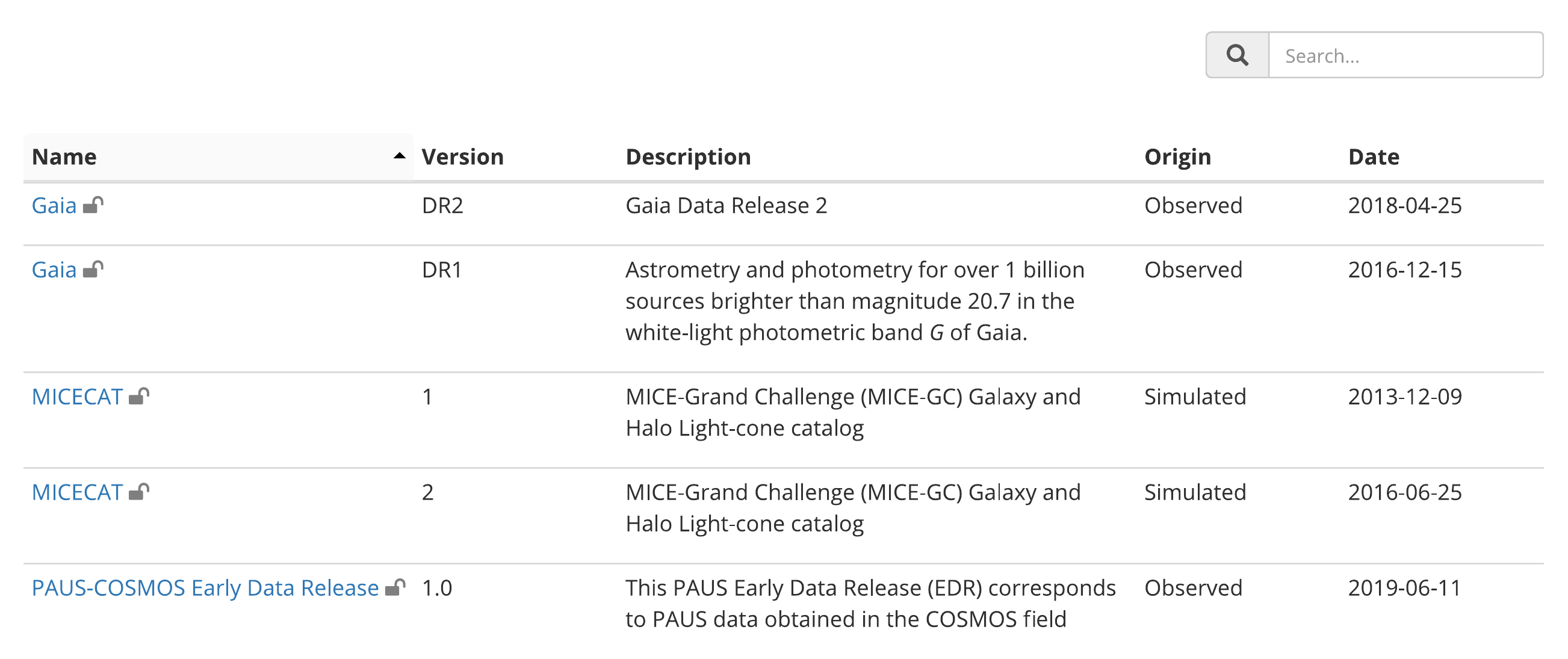}
\caption{
Catalog selection page, showing a subset of current public catalogs.
}
\label{fig:page catalogs}
\end{figure*}

Authentication is required in order to access CosmoHub's main features. For this, users have to enter their credentials in the login form. If they forgot their credentials, they can enter their email address in the reset password form. A link will be sent to that address that allows them to set up a new password.

If they do not have an account, they can fill in a registration form. A captcha\footnote{\url{https://www.google.com/recaptcha}} protects this form from automated/spam registration attempts. Upon completion, users receive an email with a link they must open to confirm ownership of the email address provided and, immediately afterwards, they are granted access to the public catalogs. Access to private data from additional projects has to be manually validated by project administrators. Requests are usually processed within the day, using a specific interface (not shown here).

Finally, users can access their profile page which displays their personal details such as name, email address and their project membership. They can update this information, including their password, or request the removal of their account.
All personal data from registered users is stored and processed following GDPR regulations.

\subsubsection*{Interactive exploration}

Just after logging in, users are presented with the catalog selection page.
As shown in Figure \ref{fig:page catalogs}, each catalog appears listed with a name, a short description, the origin (observed or simulated data) and the date it was uploaded to CosmoHub. The headers on the top of each column allow to sort the list for each field, while the top search box allows user to restrict the listing to only those catalogs

\begin{figure*}
\centering
\includegraphics[width=1\linewidth,frame]{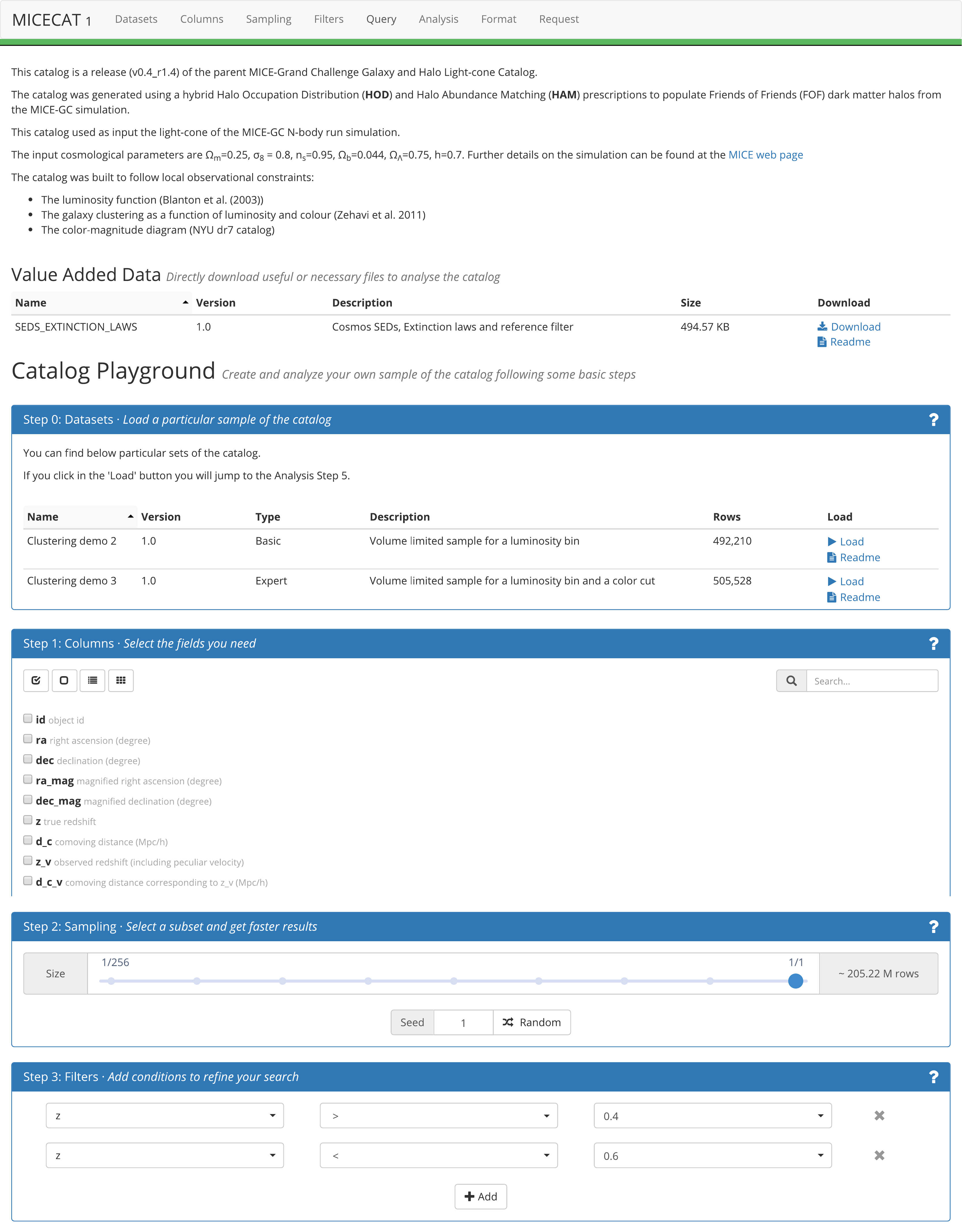}
\caption{
Catalog page upper half, showing catalog's description, valued added data and steps 0 to 3.
}
\label{fig:page query1}
\end{figure*}

\begin{figure*}
\centering
\includegraphics[width=1\linewidth,frame]{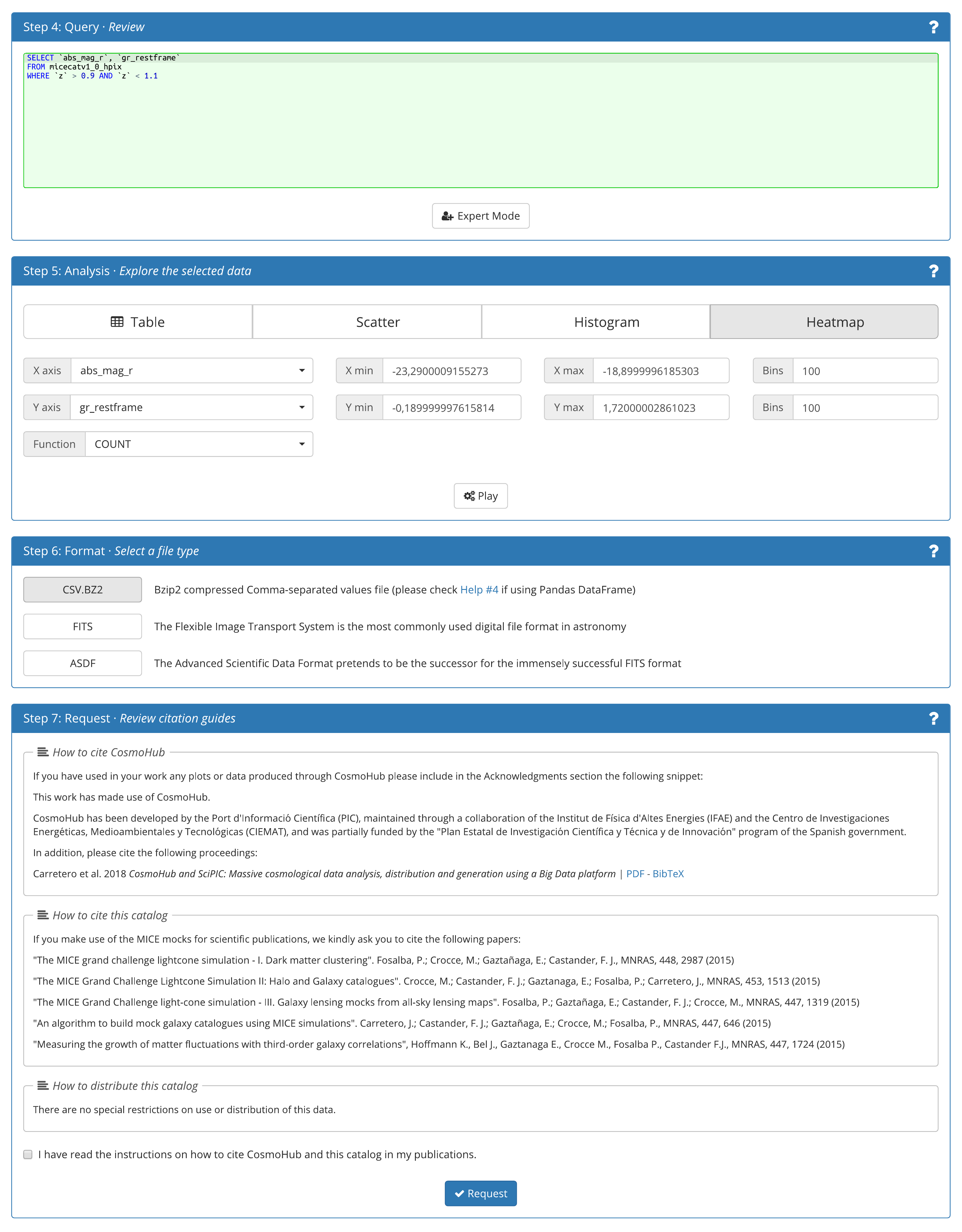}
\caption{
Catalog page bottom half, showing steps 4 to 7.
}
\label{fig:page query2}
\end{figure*}

Selecting any entry brings them to the catalog page (Figures \ref{fig:page query1} and \ref{fig:page query2}), where they can build their own custom catalogs or subsets for interactive exploration and/or download.
In order to guide them, the subset construction process is divided in a series of steps, which can be traversed through scrolling or using the navigation bar fixed at the top.

The first piece of information a user encounters is a complete description of the catalog, usually provided by the catalog owner. Just below, an optional section called "Value Added Data" contains links and documentation to additional data that complements this catalog and that may be useful to analyze it, such as filter curves or extinction maps.

Users can use a predefined dataset (Step 0) or create a new one from scratch (Step 1). Predefined datasets are curated options with specific purposes, although users can modify them to suit their needs.
For each dataset listed in Step 0, its name, version, description, type and number of rows is shown.
There are two types of predefined datasets: \textit{basic} datasets use the guided interface to configure the subset, while \textit{expert} datasets resort to setting up directly the SQL statement in expert mode.
In order to build a custom catalog from scratch, users start choosing the set of columns they need from Step 1. The search box on the top right allows users to filter the columns display looking for partial matches on names and comments.

Steps 2 and 3 represent two methods available to users for restricting the number of rows contained in their custom catalog. On the one hand, with row sampling (Step 2), Hive can be configured to only read a fraction of the files that store the catalog's data. In an attempt to deliver statistically unbiased subsets, rows are divided in those files not following any actual property \textemdash{}such as position, mass, luminosity, etc.\textemdash{} but a pseudo-random value, usually a surrogate key. In addition, users may specify in Step 3 any arbitrary criteria to further restrict the resulting rows. A filtering criteria consists of a column, an operator and a constant value. If multiple criteria are specified its effects are combined, thus only rows fulfilling all criteria are returned.

Step 4 displays the corresponding SQL statement constructed from the options selected in the previous steps. If the guided interface capabilities are not enough, or users are proficient using SQL, the \textit{Expert} mode can be enabled by clicking a button. From that point on, the guided interface will be disabled and the SQL sentence can only be modified by manually editing it in the text area.

\begin{figure}[t!]
\centering
\includegraphics[width=1\linewidth]{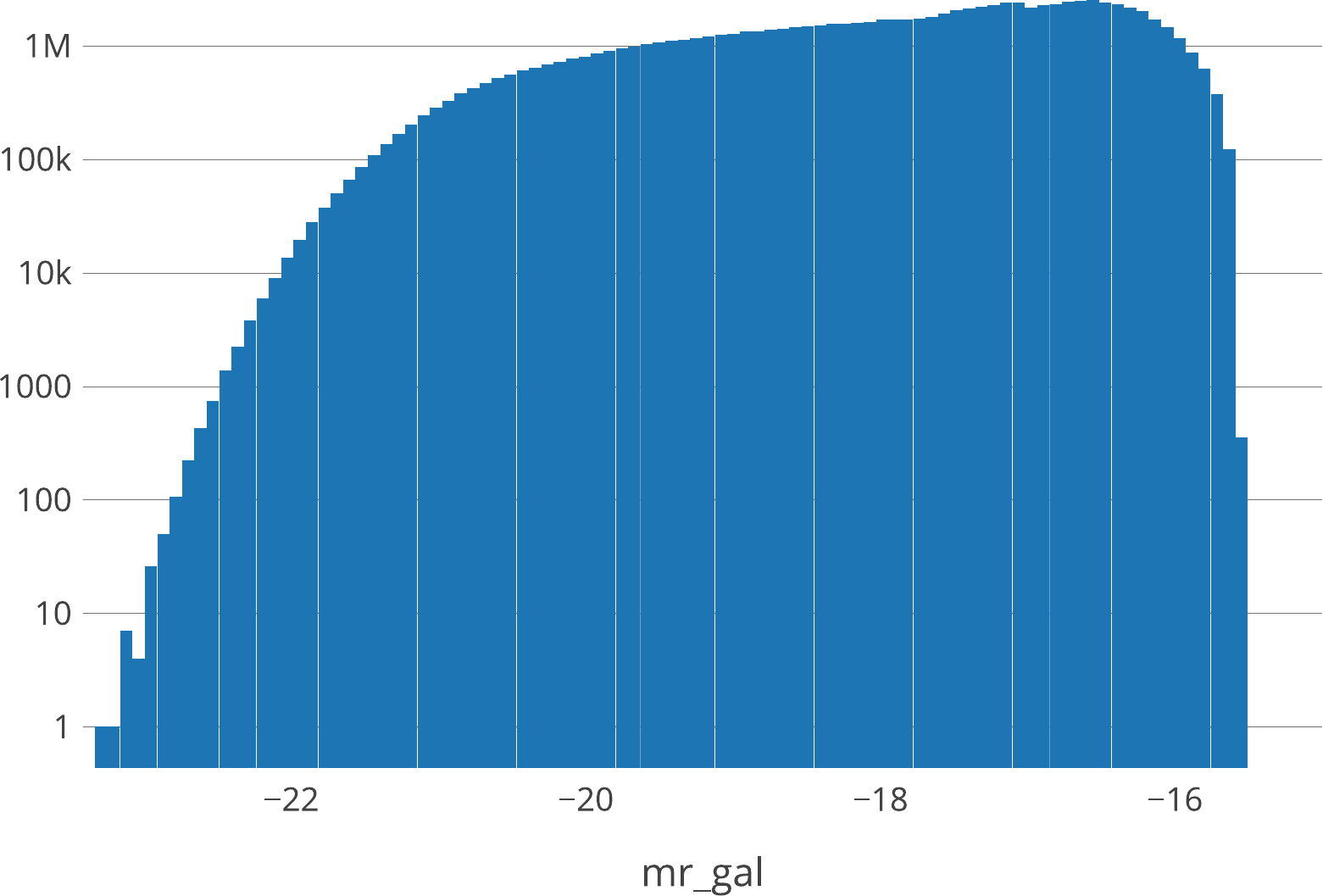}
\caption{
1D histogram, displaying the number of galaxies in MICECAT2 with true redshift between 0.4 and 0.6, grouped by absolute magnitude in a hundred uniformly sized bins.
}
\label{fig:page hist}
\end{figure}

\begin{figure}[t!]
\centering
\includegraphics[width=1\linewidth]{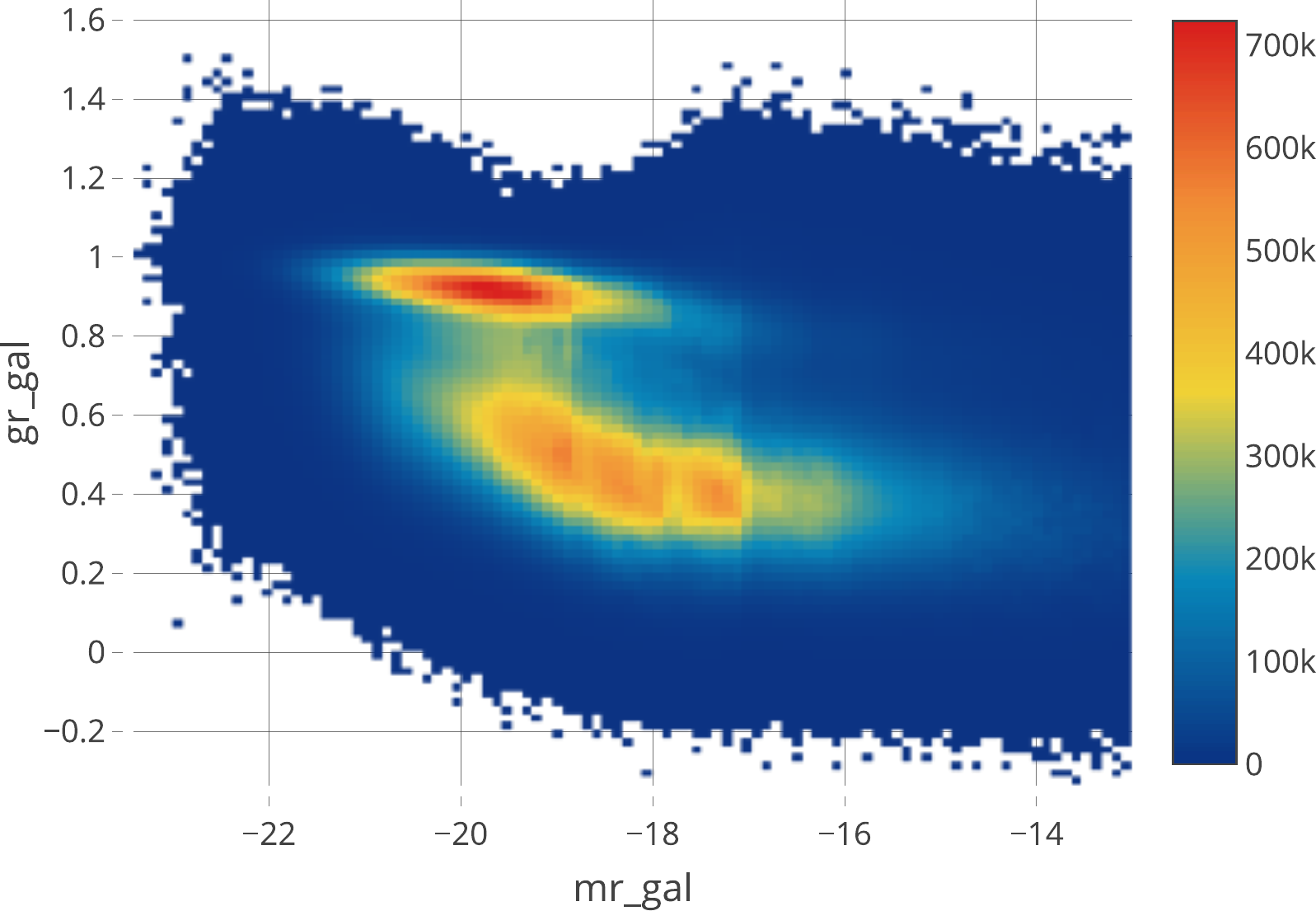}
\caption{
2D heatmap, displaying a color-magnitude diagram of MICECAT2.
}
\label{fig:page heatmap}
\end{figure}

Once the custom catalog has been defined, users can interactively explore its properties using any of the 4 visualization tools in Step 5. (i) The table preview shows the first 20 rows in the subset and it is mainly used to have a glance at the results. (ii) The scatter plot allows to display the relationship between several properties. However, as it cannot aggregate data, it is limited to ten thousand points. (iii) 1D histograms can be generated from any column using a configurable number of uniformly sized bins. The upper and lower bound of the bins are automatically hinted based on the column statistics stored by Hive. An example is shown in Figure \ref{fig:page hist}. (iv) 2D heatmaps, as scatter plots, display the relation between two properties using rectangular bins. As with 1D histograms, bin ranges are filled in from column statistics. Also, the metric can be selected between \textit{COUNT}, \textit{AVG}, \textit{MAX} or \textit{MIN}, in order to display the number of rows, average, maximum or minimum value for each bin, respectively. Figure \ref{fig:page heatmap} shows an example.

After filling in the required fields, pressing the \textit{Play} button will start the process to generate the visualization. The status is presented using a progress bar with two different stages. In static color, the amount of work that has been completed, and in a moving pattern the amount that is being executed right now. When the progress bar becomes all filled and static, the results are downloaded and displayed on screen. All plots can have customized display options such as axis scaling (linear or logarithmic), axis direction (increasing or decreasing) or switching to a cumulative plot. Also, visualizations can be zoomed in and out, exported as a Portable Network Graphics (PNG) image or downloaded as a CSV file for additional processing.

\subsubsection*{Custom catalogs}

Whenever users are satisfied with the properties of their custom catalog, they can request it to be generated and stored in a specific format to be later downloaded. Step 6 shows the supported formats: CSV with BZip2 compression, FITS and ASDF. Finally, after selecting the desired download format, users have to read and accept the corresponding data usage and citation guidelines in Step 7. Once requested, the custom catalog is assigned a unique identifier, so that users may track its progress

\begin{figure*}
\centering
\includegraphics[width=1\linewidth,frame]{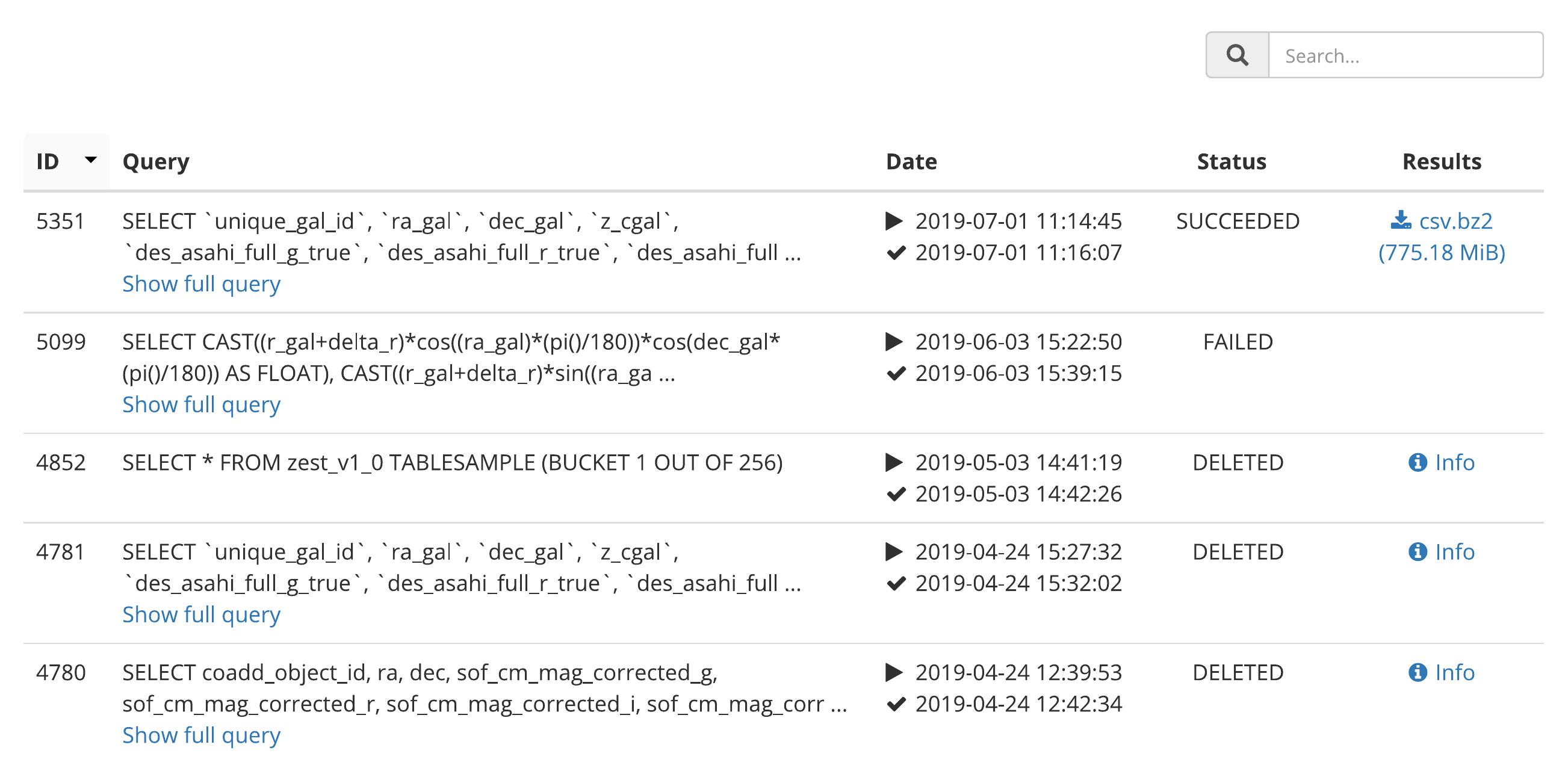}
\caption{
Activity page, used to follow custom catalog creation progress and to download them when they are ready.
}
\label{fig:page activity}
\end{figure*}

When a custom catalog is completed, users receive an email directing them to the Activity page (see Figure \ref{fig:page activity}) in order to download it. Users may also use this page to follow the progress of their custom catalog requests. Finished catalogs are kept a minimum of 30 days and are eventually deleted to maintain enough free storage space.

\section{Results}
\label{S:Results}

In this section we present some of the most relevant results and successful use cases of the version of CosmoHub as described in this article, which was commissioned in October 2016. The results are separated into two sections, the first one contains a quantitative analysis of the volume, timing and performance of the service as a whole, while the second section contains specific scientific applications where CosmoHub is being used.

\subsection{Quantitative analysis}
\label{S:Quantitative analysis}

\begin{figure}
\centering
\includegraphics[width=1.0\columnwidth]{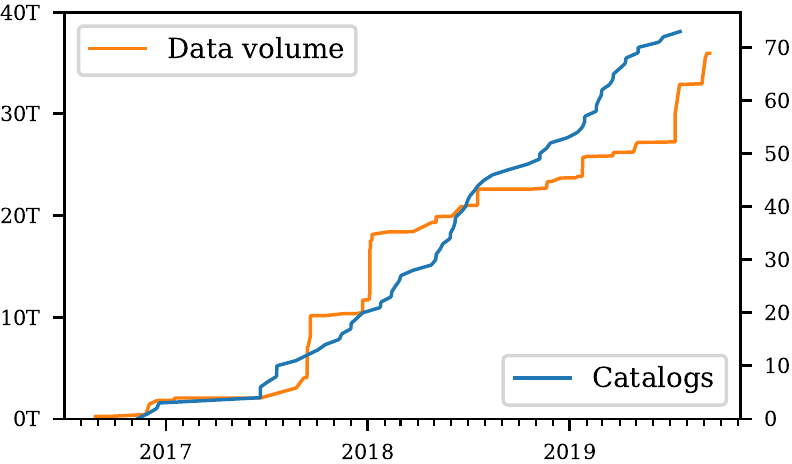}
\caption{Evolution over time of the number of available catalogs and size of published data.}
\label{fig:volume grow}
\end{figure}

Since this version opened for public use, CosmoHub has been constantly growing in all relevant metrics, such as number of users, number of catalogs and volume of published data (see Figure \ref{fig:volume grow}). Note that the information in this figure only accounts for data published through CosmoHub, excluding custom catalogs.

Since late 2016, more than 600 new users have opened an account, and more than 4,000 custom catalogs and nearly 10,000 interactive queries have been delivered. Data growth has been limited by the available storage space. By the end of 2019, our current expectations are to double the amount of published data.

\begin{figure}
\centering
\includegraphics[width=1.0\columnwidth]{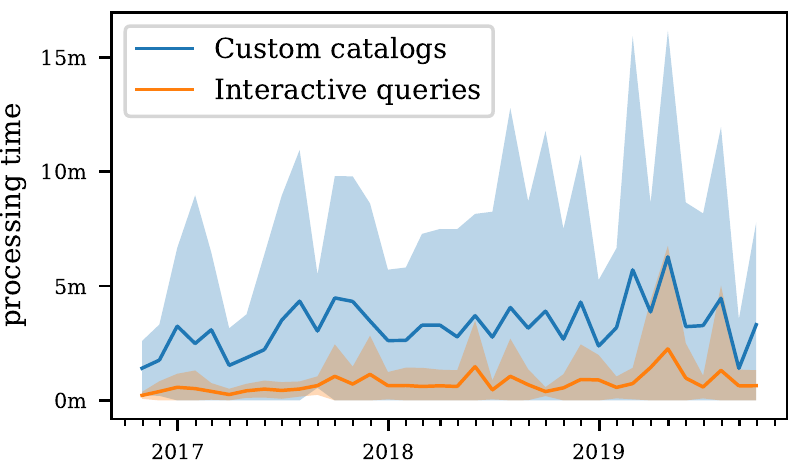}
\caption{Processing time in minutes (monthly average) for batch catalogs and interactive queries. Shaded area shows one sigma deviation.}
\label{fig:average time}
\end{figure}

At the same time, performance in terms of response time has been stable. Figure \ref{fig:average time} shows that the average response time for the execution of interactive queries and the generation of custom catalogs has barely increased over time. Statistical fluctuations are due to resource contention with concurrent queries and scheduling overheads, among others.

It is worth noticing that, since its commissioning, the architecture and configuration of the Hadoop platform has seen several reorganizations, although the only resource that has been increased is the storage capacity. Therefore, by improving and tuning the platform, we have been able to cope with the growth in users and data volume and to keep the response time stable. Nevertheless, we do not expect to be able to squeeze much more performance of the actual setup so, in the future, additional resources will be required in order to reduce or keep the response times stable.

\begin{figure}
\centering
\includegraphics[width=1.0\columnwidth]{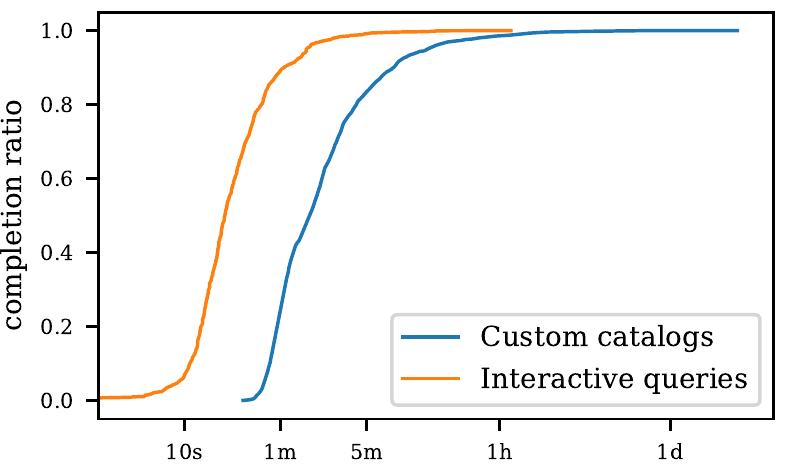}
\caption{Completion ratio as a factor of processing time for batch catalogs and interactive queries.}
\label{fig:completion ratio}
\end{figure}

Figure \ref{fig:completion ratio} provides information about the distribution of response times by plotting the completion ration for both interactive queries and custom catalogs. The completion ration is defined as the fraction of queries completed after a given elapsed time. Note that the orchestration of the different tasks on the cluster nodes has a minimum overhead of about 10-12 seconds. Only queries that can be answered directly from statistics, such as selecting the number of rows or the maximum value of a column without filters, return in a shorter time.

About 66.9\% of all interactive queries finish in less than 30 seconds, while 96.8\% of them finish in less than 2 minutes. These response times enable users to interactive explore any dataset, without worrying about its volume or the complexity of the query.
On the other hand, for custom catalogs, 71.0\% are produced in less than 3 minutes, while 97.4\% of them finish in 30 minutes. Although not as relevant as for interactive queries, keeping a low response time for the generation of custom catalogs is also important to keep a good user experience.

\begin{figure}
\centering
\includegraphics[width=1.0\columnwidth]{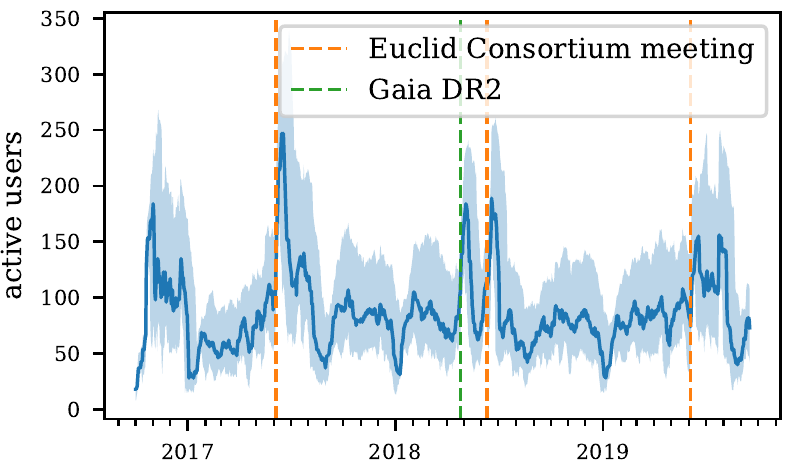}
\caption{Active users over time, with relevant milestones highlighted. Upper and lower bounds correspond to active users within a window of 28 days and 7 days respectively, while middle line is for a 14 days window.}
\label{fig:active users}
\end{figure}

Figure \ref{fig:active users} shows the evolution of active users over time. Dates of particular scientific events are overlayed. A clear correlation can be seen, particularly with the Euclid Consortium meeting held yearly every June, where interest on newly released data boosts user activity. A peak in activity can also be seen coinciding with Gaia Data Release 2 (DR2\footnote{\url{https://www.cosmos.esa.int/web/gaia/dr2}}). The catalog was mirrored in CosmoHub in less than 12h and we were able to release it almost simultaneously to the official announcement at the Europen Space Astronomy Centre (ESAC\footnote{\url{http://www.esa.int/About_Us/ESAC}}).

\subsection{Scientific applications}
\label{S:Scientific applications}

\begin{table}
\centering
\begin{tabularx}{\columnwidth}{|X r r r|} 
  \hline
  Catalog & Rows & Fields & Size \\ 
  \hline \hline
  ALHAMBRA & & & \\
   --- S/G classified & 422 K & 7 & 10.4 MiB \\
   --- photometric redshifts & 441 K & 113 & 149.5 MiB \\
  CFHTLens & 29 M & 129 & 7.5 GiB \\
  COSMOS 2015 & 1.2 M & 537 & 1.6 GiB \\
  Gaia DR1 & 1,143 M & 62 & 174.9 GiB \\
  Gaia DR2 & 1,693 M & 98 & 486.5 GiB \\
  KiDS DR4 & 100 M & 306 & 89.6 GiB \\
  MICECAT 1 & 205 M & 91 & 59.9 GiB \\
  MICECAT 2 & 500 M & 122 & 211.1 GiB \\
  PAUS-COSMOS EDR & 6.5 K & 126 & 2.8 MiB \\
  PAU-MillGas Lightcone & 7.4 M & 34 & 9.0 GiB \\
  Zest & 132 K & 71 & 20.9 MiB \\
  \hline
\end{tabularx}
\caption{
List of public catalogs in CosmoHub.
}
\label{table:catalogs}
\end{table}

CosmoHub supports multiple international cosmology projects, 
the most relevant in terms of users and data volume being Euclid
\footnote{\url{http://sci.esa.int/euclid/}}.
Euclid simulations require the production of extremely large datasets. In their Flagship simulation\footnote{\url{https://www.euclid-ec.org/?page_id=4133}} (paper in preparation), with an estimated final size of 20 TiB and nearly $30 \times 10^9$ objects, they are using as input the largest dark-matter halo catalog up to date, with 5.5 TiB and $40 \times 10^9$ objects. CosmoHub is currently providing access to the entire halo catalog and to several productions of mock galaxy catalogs (these account for about half of the data stored in CosmoHub). The Euclid cosmological simulations validation team makes heavy use of the exploration capabilities of CosmoHub in order to validate data prior to its release to the full collaboration. Then, Euclid scientists use CosmoHub to download customized subsets of the data for further analysis and processing.

At the time of this writing, up to 24 publications have acknowledged CosmoHub contribution to their results. Projects such as PAU Survey\footnote{\url{https://www.pausurvey.org/}} and MICE\footnote{\url{http://maia.ice.cat/mice/}} use CosmoHub as the official and primary channel for the distribution of their data (\cite{10.1093/mnras/stz204} and \cite{10.1093/mnras/sty3129}). Other projects such as DES\footnote{\url{https://www.darkenergysurvey.org/}} and Gaia\footnote{\url{http://sci.esa.int/gaia/}} have a replica of their most important releases, and several publications have made use of them (see \cite{blackbody} and \cite{2018MNRAS.481.5451S}). The subset of public catalogs in CosmoHub (as of October 2019) is shown in Table \ref{table:catalogs}.

Finally, other applications not based solely on the distribution of data are also present, such as the one from the DES clustering science working group. They made intensive use of the exploration capabilities to define and test many different arbitrary galaxy subsamples to estimate the Baryon Acoustic Oscillations (BAO) feature in the galaxy distribution \citep{2019MNRAS.482.2807C}. 
Some particularly useful applications are shown in \ref{A:Useful applications}. Lastly, some projects have also cited CosmoHub as a state-of-the-art reference to their own data publication procedures, such as \cite{haac} and \cite{IllustrisTNG}.

\section{Conclusions and future work}
\label{S:Conclusions}

This paper presents CosmoHub, our vision to enable the interactive exploration and distribution of large cosmological datasets on top of Hadoop.
The paper describes the main features and capabilities of CosmoHub from the user's point of view but, more importantly, it also details all the research and decisions made regarding its design and implementation.

Regarding the design (explained in section \ref{S:Solution design}), it is focused on satisfying a set of needs (enumerated in section \ref{S:Objectives}) gathered from the scientific community, while the experience gained through the early prototypes (described in section \ref{S:Early prototypes}) helped pave the way to achieve its current success.
In particular, the easy to use requirement has been met by implementing CosmoHub as a web application, with a guided process to remove any SQL knowledge dependency, and the support for common data formats such as CSV and FITS.
Also, the ability to produce visualizations to get insight over thousands of millions of rows in just a few seconds fulfills the interactive exploration requirement.
Several projects such as PAUS, MICE and the Euclid simulation group have selected CosmoHub as their primary data distribution service, which was also one of our objectives.

The decision to delegate CosmoHub's data processing to Hadoop and Hive (see section \ref{S:Hadoop}) has proven to be a good choice.
The reliability and high performance exhibited by this data warehouse solution, combined with the usability and powerful features implemented in CosmoHub's interface, have enabled to steadily support a growing number of users and projects.
Also, the great scalability of the platform has allowed to keep response times low at all times (see section \ref{S:Quantitative analysis}), in spite of the constant increase in data volume.
In the end, the results presented support the fact that CosmoHub is providing a useful service to the scientific community with a high quality of service, as proven by the use of CosmoHub by some of the most relevant projects in cosmology (see section \ref{S:Scientific applications}).

When CosmoHub entered into service in late 2016, it was the first project to apply Hadoop to the analysis and distribution of large cosmological datasets. Over these years we have learnt a lot from both our own experience and user's feedback. In fact, we are already working on the next iteration of CosmoHub which will include a lot of improvements based on this experience:

(i) Regarding the Hadoop platform, upgrade it to the latest HDP release. The most exciting new features include the possibility to reduce replica overhead using erasure coding, the ability to access a read-only view of externally provided storage and the implementation of materialized views in Hive to speed up join queries.

(ii) From CosmoHub application's perspective, add the ability for users to upload their own catalogs and to publish and share them with other users, extend and optimize the visualization tools performance using a binary protocol, and improve the general responsiveness of the user interface, among many other new features. 

With all this future work under way, we are prepared to keep pushing forward and to help put in place the next generation of services for managing large volumes of structured scientific data.

\section*{Acknowledgements}

CosmoHub has been partially funded through projects of the Spanish national program ``Programa Estatal de I+D+i'' of the Spanish government.
The support of the ERDF fund is gratefully acknowledged.

We are also deeply grateful to all PIC's staff, both for the encouragement from the management and for the tremendous help and support we get from our IT colleagues.

Finally, our most sincere thanks to CosmoHub's community of users, for their invaluable feedback and advice.








\section*{References}

\bibliographystyle{model2-names}

\bibliography{CosmoHub.bib}

\onecolumn
\appendix
\section{Particularly useful applications}
\label{A:Useful applications}

This appendix describes in detail several representative use cases that make use of the custom catalog generation capabilities in CosmoHub.
Each application includes the full SQL statement that was used, along with the time it took to complete. The timings measured in this section, unlike the results in section \ref{S:Results}, were performed having exclusive use of the entire Hadoop platform.

\subsection*{MICECAT1 clustering sample}

In this application, we want to generate a subset of MICECAT1 in order to compute the projected 2-point correlation function on it. Thus, we selected the right ascension, the declination and the comoving distance columns, and we filtered on a redshift shell (with \texttt{z} between 0.3 and 0.4) and also on a magnitude range (with absolute magnitude on the \texttt{r} band between -22 and -21). The creation of this custom catalog takes only 14 seconds and generates a CSV.BZ2 file of 4.99 MiB containing 492,210 rows.

\begin{verbatim}
SELECT `ra`, `dec`, `d_c`
FROM micecat_v1
WHERE `z` > 0.3 AND `z` < 0.4
  AND `abs_mag_r` < -21 AND `abs_mag_r` > -22 
\end{verbatim}

\subsection*{DES Y1A1 BAO main sample}

Another interesting application was the generation of the BAO sample for DES.
This sample, described in section 3 of \citet{2019MNRAS.482.2807C}, is a subset of DES Y1 data that, according to the article, represents ``red galaxies with a good compromise of photo-z accuracy and number density, optimal for the BAO measurement''. The query below implements the criteria shown in Table 1 of the paper.
The fast response times of CosmoHub were particularly useful to interactively refine the parameters of the sample, which is now available also as a predefined dataset within one of the DES private catalogs. An interactive query to visualize the number of objects as a function of the photometric redshift takes 31 seconds. Exporting the sample into a CSV.BZ2 file of 55.6 MiB containing 2.7 million objects takes 39 seconds.

\begin{verbatim}
SELECT coadd_objects_id, ra, dec, mean_z_bpz_hiz, z_mc_bpz_hiz, t_b_hiz, odds_hiz
FROM des_y1
WHERE (mag_auto_i > 17.5) AND (mag_auto_i < 22)
  AND (mag_auto_i < 19.0 + 3.0*mean_z_bpz_hiz)
  AND (ra < 15 or ra > 290 or dec < -35)
  AND (flags_badregion <= 3 and flags_gold = 0)
  AND (spread_model_i + (5.0/3.0)*spreaderr_model_i > 0.007)
  AND ((mag_auto_i - mag_auto_z) + 2.0*(mag_auto_r - mag_auto_i) > 1.7)
  AND ((mag_auto_g - mag_auto_r) BETWEEN -1. and 3.)
  AND ((mag_auto_r - mag_auto_i) BETWEEN -1. and 2.5)
  AND ((mag_auto_i - mag_auto_z) BETWEEN -1. and 2.)
  
\end{verbatim}

\subsection*{GAIA DR2 pseudo Healpix map}

The custom catalog feature can also be used in conjunction with the FITS format to create HEALPiX maps. For instance, in this application the following query was used to create a partial map with explicit indexing\footnote{See \url{https://healpix.sourceforge.io/data/examples/healpix_fits_specs.pdf} for more information about how HEALPiX data is stored as FITS.} estimating the average of the \textit{Standard error of parallax (Angle[mas])} for each pixel. The pixel identifier is taken from the \texttt{\_hpix\_12\_nest} column. The generation of this custom catalog produces a FITS file of 156 million rows and 1.74 GiB in size in 65 seconds.

\begin{verbatim}
SELECT `_hpix_12_nest`, AVG(parallax_error)
FROM gaia_dr2 
GROUP BY `_hpix_12_nest`
\end{verbatim}

\subsection*{Euclid True Universe FITS file}

CosmoHub stores and distributes a large amount of data for the Euclid Cosmological Simulations Working Group (CSWG) which is then used as input for different image simulator pipelines. This application uses a complex SQL statement to generate an individual input from the catalogs stored in CosmoHub. The resulting FITS file has the correct format, the proper field names and the correct units. This allows Euclid scientists to easily test their codes on a smaller scale, while at the same time enabling them to iterate and provide feedback much faster. In this particular example, the FITS file generated contains 1823344 objects, occupies 406.9 MiB and was produced in 22s.

\begin{verbatim}
SELECT CAST(((gal.halo_id * 10000) + gal.galaxy_id) AS bigint) AS SOURCE_ID, 
  CAST(gal.ra_gal AS float) AS RA, 
  CAST(gal.dec_gal AS float) AS DEC, 
  CAST(gal.ra_gal_mag AS float) AS RA_MAG, 
  CAST(gal.dec_gal_mag AS float) AS DEC_MAG, 
  CAST(gal.observed_redshift_gal AS float) AS Z_OBS, 
  CAST(gal.abs_mag_r01_evolved AS float) AS TU_MAG_R01_SDSS_ABS, 
  CAST(-2.5*log10(gal.sdss_r01) - 48.6 AS float) AS TU_MAG_R01_SDSS, 
  CAST(gal.sed_cosmos AS float) AS SED_TEMPLATE, 
  CAST(ROUND(gal.ext_curve_cosmos) AS smallint) AS EXT_LAW, 
  CAST(gal.ebv_cosmos AS float) AS EBV, 
  CAST(gal.logf_halpha_model3_ext AS float) AS HALPHA_LOGFLAM_EXT, 
  CAST(gal.logf_hbeta_model3_ext AS float) AS HBETA_LOGFLAM_EXT, 
  CAST(gal.logf_o2_model3_ext AS float) AS O2_LOGFLAM_EXT, 
  CAST(gal.logf_o3_model3_ext AS float) AS O3_LOGFLAM_EXT, 
  CAST(gal.logf_n2_model3_ext AS float) AS N2_LOGFLAM_EXT, 
  CAST(gal.logf_s2_model3_ext AS float) AS S2_LOGFLAM_EXT, 
  CAST(gal.bulge_fraction AS float) AS BULGE_FRACTION, 
  CAST(gal.bulge_length AS float) AS BULGE_LENGTH, 
  CAST(gal.disk_length AS float) AS DISK_LENGTH, 
  CAST(gal.disk_axis_ratio AS float) AS DISK_AXIS_RATIO, 
  CAST(gal.disk_angle AS float) AS DISK_ANGLE, 
  CAST(gal.kappa AS float) AS KAPPA, 
  CAST(gal.gamma1 AS float) AS GAMMA1,
  CAST(gal.gamma2 AS float) AS GAMMA2, 
  CAST(gal.mw_extinction AS float) AS AV, 
  CAST(gal.euclid_vis_el_model3_odonnell_ext*1.e23 AS float) AS TU_FNU_VIS, 
  CAST(gal.euclid_nisp_y_el_model3_odonnell_ext*1.e23 AS float) AS TU_FNU_Y_NISP, 
  CAST(gal.euclid_nisp_j_el_model3_odonnell_ext*1.e23 AS float) AS TU_FNU_J_NISP, 
  CAST(gal.euclid_nisp_h_el_model3_odonnell_ext*1.e23 AS float) AS TU_FNU_H_NISP, 
  CAST(gal.blanco_decam_g_el_model3_odonnell_ext*1.e23 AS float) AS TU_FNU_G_DECAM, 
  CAST(gal.blanco_decam_r_el_model3_odonnell_ext*1.e23 AS float) AS TU_FNU_R_DECAM, 
  CAST(gal.blanco_decam_i_el_model3_odonnell_ext*1.e23 AS float) AS TU_FNU_I_DECAM, 
  CAST(gal.blanco_decam_z_el_model3_odonnell_ext*1.e23 AS float) AS TU_FNU_Z_DECAM, 
  CAST(gal.cfht_megacam_u_el_model3_odonnell_ext*1.e23 AS float) AS TU_FNU_U_MEGACAM, 
  CAST(gal.cfht_megacam_r_el_model3_odonnell_ext*1.e23 AS float) AS TU_FNU_R_MEGACAM, 
  CAST(gal.jst_jpcam_g_el_model3_odonnell_ext*1.e23 AS float) AS TU_FNU_G_JPCAM, 
  CAST(gal.pan_starrs_i_el_model3_odonnell_ext*1.e23 AS float) AS   TU_FNU_I_PANSTARRS, 
  CAST(gal.pan_starrs_z_el_model3_odonnell_ext*1.e23 AS float) AS TU_FNU_Z_PANSTARRS, 
  CAST(gal.subaru_hsc_z_el_model3_odonnell_ext*1.e23 AS float) AS TU_FNU_Z_HSC, 
  CAST(gal.gaia_g_el_model3_odonnell_ext*1.e23 AS float) AS TU_FNU_G_GAIA, 
  CAST(gal.gaia_bp_el_model3_odonnell_ext*1.e23 AS float) AS TU_FNU_BP_GAIA, 
  CAST(gal.gaia_rp_el_model3_odonnell_ext*1.e23 AS float) AS TU_FNU_RP_GAIA, 
  CAST(gal.lsst_u_el_model3_odonnell_ext*1.e23 AS float) AS TU_FNU_U_LSST, 
  CAST(gal.lsst_g_el_model3_odonnell_ext*1.e23 AS float) AS TU_FNU_G_LSST, 
  CAST(gal.lsst_r_el_model3_odonnell_ext*1.e23 AS float) AS TU_FNU_R_LSST, 
  CAST(gal.lsst_i_el_model3_odonnell_ext*1.e23 AS float) AS TU_FNU_I_LSST, 
  CAST(gal.lsst_z_el_model3_odonnell_ext*1.e23 AS float) AS TU_FNU_Z_LSST, 
  CAST(gal.lsst_y_el_model3_odonnell_ext*1.e23 AS float) AS TU_FNU_Y_LSST, 
  CAST(gal.kids_u_el_model3_odonnell_ext*1.e23 AS float) AS TU_FNU_U_KIDS, 
  CAST(gal.kids_g_el_model3_odonnell_ext*1.e23 AS float) AS TU_FNU_G_KIDS, 
  CAST(gal.kids_r_el_model3_odonnell_ext*1.e23 AS float) AS TU_FNU_R_KIDS, 
  CAST(gal.kids_i_el_model3_odonnell_ext*1.e23 AS float) AS TU_FNU_I_KIDS, 
  CAST(gal.2mass_j_el_model3_odonnell_ext*1.e23 AS float) AS TU_FNU_J_2MASS, 
  CAST(gal.2mass_h_el_model3_odonnell_ext*1.e23 AS float) AS TU_FNU_H_2MASS, 
  CAST(gal.2mass_ks_el_model3_odonnell_ext*1.e23 AS float) AS TU_FNU_KS_2MASS, 
  CAST(SHIFTRIGHT(gal.hpix_29_nest, (29-5)*2) AS bigint) AS hpix_5_nest 
FROM cosmohub.flagship_mock_sc456 AS gal 
WHERE SHIFTRIGHT(hpix_29_nest, (29-5)*2) = 7155
\end{verbatim}

\end{document}